\documentclass[11pt,letterpaper]{article}

\usepackage{sectsty}
\usepackage{notoccite}
\usepackage{graphics} 
\usepackage{amsmath} 
\usepackage{amssymb}  
\usepackage{color,xspace}
\usepackage[hidelinks]{hyperref}
\usepackage{graphicx}
\usepackage{subcaption}
\usepackage{cite}
\usepackage{cleveref}
\usepackage{bm}
\usepackage{bbm}
\usepackage{tikz}
\usepackage{epstopdf} 
\usepackage{ntheorem}
\usepackage{enumitem}
\usepackage{mdframed}   
\usepackage{sectsty}

\usepackage{subcaption}
\usepackage[margin=1.0in]{geometry}

\usepackage{caption}

\allowdisplaybreaks

\newcommand{\indinf}{{\infty \rightarrow \infty}}
\newcommand{\DDelta}{\mathbf{\Delta}}
\newcommand{\xx}{\mathbf{x}}
\newcommand{\uu}{\mathbf{u}}
\newcommand{\ww}{\mathbf{w}}
\newcommand{\bfeta}{\boldsymbol{\eta}}

\newcommand{\sA}{\mathbf{A}}
\newcommand{\sB}{\mathbf{B}}
\newcommand{\sR}{\mathcal{R}}
\newcommand{\sQ}{\mathcal{Q}}

\newcommand{\Phix}{\mathbf{\Phi}_x}
\newcommand{\Phiu}{\mathbf{\Phi}_u}
\newcommand{\KK}{\mathbf{K}}
\newcommand{\PPhi}{\mathbf{\Phi}}
\newcommand{\tildePhix}{\widetilde{\mathbf{\Phi}}_x}
\newcommand{\tildePhiu}{\widetilde{\mathbf{\Phi}}_u}
\newcommand{\tildeww}{\widetilde{\mathbf{w}}}

\newcommand{\SLSOCP}{\text{SLSOCP}}

\newcommand{\Phixk}{\mathbf{\Phi}_{x,k}}
\newcommand{\Phiuk}{\mathbf{\Phi}_{u,k}}
\newcommand{\Phixkp}{\mathbf{\Phi}_{x,k+1}}
\newcommand{\Phiukp}{\mathbf{\Phi}_{u,k+1}}
\newcommand{\tildePhixk}{\widetilde{\mathbf{\Phi}}_{x,k}}
\newcommand{\tildePhiuk}{\widetilde{\mathbf{\Phi}}_{u,k}}
\newcommand{\tildePhixkp}{\widetilde{\mathbf{\Phi}}_{x,k+1}}
\newcommand{\tildePhiukp}{\widetilde{\mathbf{\Phi}}_{u,k+1}}

\newcommand{\augdist}{lumped uncertainty}

\newcommand{\augdists}{lumped uncertainties}
\newcommand{\augslsmpc}{lumped-SLS-MPC}

\newtheorem{theorem}{Theorem}
\newtheorem{lemma}{Lemma}    

\newtheorem{remark}{Remark}

    \newtheorem*{proof}{Proof}    
\newtheorem{assumption}{Assumption}    
\newtheorem{definition}{Definition}    
\newtheorem{problem}{Problem}    

\DeclareMathOperator*{\argmin}{arg\,min}

\title{{\LARGE \bf System Level Synthesis-based Robust Model Predictive Control through Convex Inner Approximation}}

\author{Shaoru Chen, Nikolai Matni, Manfred Morari, Victor M. Preciado
	\thanks{The authors are with the Department of Electrical and Systems Engineering, University of Pennsylvania. Email: \{srchen, nmatni, morari, preciado\}@seas.upenn.edu. Our codes are publicly available at \url{https://github.com/ShaoruChen/Lumped-Uncertainty-SLS-MPC}.
	}
}

\date{}

\begin{document}
\pagestyle{plain}
\maketitle

\begin{abstract}

We propose a robust model predictive control (MPC) method for discrete-time linear time-invariant systems with norm-bounded additive disturbances and model uncertainty. In our method, at each time step we solve a finite time robust optimal control problem (OCP) which jointly searches over robust linear state feedback controllers and bounds the deviation of the system states from the nominal predicted trajectory. By leveraging the System Level Synthesis (SLS) framework, the proposed robust OCP is formulated as a convex quadratic program in the space of closed-loop system responses. When an adaptive horizon strategy is used, we prove the recursive feasibility of the proposed MPC controller and input-to-state stability of the origin for the closed-loop system. We demonstrate through numerical examples that the proposed method considerably reduces conservatism when compared with existing SLS-based and tube-based robust MPC methods, while also enjoying low computational complexity.

\end{abstract}

\section{Introduction}
In model predictive control (MPC), a finite time constrained optimal control problem (OCP) is solved at each time step and the first optimal control input is applied. When the system model is uncertain, robust model predictive control explicitly takes the system uncertainty into account by solving a robust OCP at each time step to guarantee that the state and control input constraints are robustly satisfied for the closed-loop system. Although in theory dynamic programming~\cite[Chapter 15]{borrelli2017predictive} can exactly solve the resulting robust OCPs, such methods suffer from prohibitive computational complexity, making them impractical. This has motivated the development of alternative solutions to the robust OCPs which aim to reach a reasonable compromise between conservatism as measured by the size of the set of feasible initial states, and computational complexity. 


For linear time-invariant (LTI) systems with only uncertain additive disturbances, various closed-loop methods have been proposed to solve the robust OCPs with feedback policies as decision variables \cite{kouvaritakis2000efficient, goulart2006optimization, mayne2005robust, langson2004robust, rakovic2012homothetic, rakovic2012parameterized, sieber2021system, lofberg2003approximations}. When model uncertainty is present, robust MPC becomes more challenging since the deviation from the nominal predicted trajectory depends on the system states and the feedback policy to be designed. For polytopic model uncertainty, robust MPC methods bsed on linear matrix inequalities (LMI) are presented in \cite{kothare1996robust, schuurmans2000robust, kouvaritakis2000efficient} where affine state feedback policies are considered. For system with both model uncertainty and additive disturbances, the tube-MPC method in~\cite{langson2004robust} designs a feedback policy that contains all possible trajectories inside a tube. In~\cite{chen2020robust, dean2019safely}, linear state feedback policies are considered and the robust OCPs are reformulated in the space of closed-loop system responses through the System Level Synthesis (SLS) framework~\cite{ANDERSON2019364}. The authors in \cite{bujarbaruah2020robust} apply a disturbance feedback policy and incorporate the policy parameters into constraint tightening of the robust OCP. In~\cite{bujarbaruah2021simple}, the authors propose to lump both model uncertainty and additive disturbance into a net-additive disturbance with larger size and calculate a disturbance feedback policy for robust MPC design to address the enlarged disturbances. 

Inspired by~\cite{bujarbaruah2021simple}, in this work we describe the uncertain system dynamics as the sum of the nominal dynamics and the \augdist{} that captures the deviation from the nominal predicted state at each time instant. When model uncertainty is present, the \augdist{} is a function of the system state and control input, and therefore of the feedback policy to be designed. Unlike~\cite{bujarbaruah2021simple}, which uses a conservative uniform over-approximation of the \augdist{} for robust MPC design, we exactly characterize the dynamics of the \augdist{} under a linear state feedback policy. This is made possible through SLS, which gives a transparent description of the interaction between the \augdist{} and the system states and control inputs in closed-loop. Based on the dynamics of the \augdist{}, our proposed robust MPC method, which we call \augdist{} SLS MPC, solves a convex inner approximation~\footnotemark of the robust OCP which jointly optimizes over a linear state feedback policy and the norm upper bounds on system state deviation. Our contributions are as follows.
\footnotetext{A convex inner approximation of a robust OCP is a convex OCP that searches over a smaller set of robust controllers compared with the original one. It usually trades off conservatism for numerical efficiency. }
\begin{enumerate}
\item We propose a novel robust MPC method, \augdist{} SLS MPC, for uncertain LTI systems with norm-bounded model uncertainty and additive disturbances. At each time step, \augdist{} SLS MPC searches over linear state feedback policies for prediction and solves a convex inner approximation of the robust OCP in the space of system responses using SLS. The convex inner approximation is derived based on the dynamics of the \augdist{} and is shown to achieve significant reduction in conservatism compared with existing SLS-based and tube-based robust MPC methods.
\item We prove the recursive feasibility and input-to-state stability (ISS) of the \augdist{} SLS MPC when an adaptive horizon strategy is used.
\item We numerically compare our proposed method with representative baseline robust MPC approaches such as tube-MPC~\cite{langson2004robust}, the SLS-based MPC method in~\cite{chen2020robust}, and the lumped-uncertainty method in~\cite{bujarbaruah2021simple}. Our proposed method is shown to outperform all baseline methods by a significant margin in conservatism as measured by the sets of feasible initial states. In addition, we show that the computational complexity of our method is better than or comparable to that of existing baselines. 
\end{enumerate}

The rest of the paper is organized as follows. The uncertain LTI system and the problem formulation are introduced in Section~\ref{sec:formulation}. With the SLS framework introduced in Section~\ref{sec:SLS}, we present the design of \augdist{} SLS MPC in Section~\ref{sec:augSLSMPC} followed by the proof of recursive feasibility (Section~\ref{sec:recursive}) and ISS (Section~\ref{sec:ISS}). A detailed comparison with baseline robust MPC methods is given in Section~\ref{sec:simulation} through numerical examples. Section~\ref{sec:conclusion} concludes the paper. 

\paragraph{Notation}
Let $\mathbb{R}_{\geq 0}$ denote the set of non-negative real numbers. For a dynamical system, we denote the system state at time $t$ by $x(t)$ and the $t$-step prediction of the state in an MPC loop by $x_t$. For two vectors $x$ and $y$, $x \leq y$ denotes element-wise comparison. For a symmetric matrix $Q$, $Q \succeq 0$ denotes that $Q$ is positive semidefinite. The notation $x_{i:j}$ is shorthand for the set $\{x_i, x_{i+1}, \cdots, x_j\}$. The norm $\lVert \cdot \rVert_\indinf$ is the $\ell_\infty$ induced norm. $S = \text{blkdiag}(S_1, \cdots, S_N)$ denotes that $S$ is a block diagonal matrix whose diagonal blocks consist of $S_1, \cdots, S_N$ arranged in the order. We represent a linear, causal operator $\mathbf{R}$ defined over a horizon of $T$ by the block-lower-triangular matrix
\begin{equation} \label{eq:BLT}
\mathbf{R} = \begin{bmatrix}
R^{0,0} & \ & \ & \ \\
R^{1,1} & R^{1,0} & \ & \ \\
\vdots & \ddots & \ddots & \ \\
R^{T,T} & \cdots &R^{T,1} & R^{T,0}
\end{bmatrix}
\end{equation} 
where $R^{i,j} \in \mathbb{R}^{p \times q}$ is a matrix of compatible dimension. The set of such matrices is denoted by $\mathcal{L}_{TV}^{T, p \times q}$ and we will drop the superscript $T, p \times q$ when it is clear from the context. Let $\mathbf{R}(i,:)$ denote the $i$-th block row of $\mathbf{R}$, and $\mathbf{R}(:,j)$ denote the $j$-th block column of $\mathbf{R}$, both indexing from $0$\footnotemark. \footnotetext{In this paper, we refer to a block matrix in a block-lower-triangular matrix $\mathbf{R}$ using its superscripts shown in Eqn.~\eqref{eq:BLT}. If we let $\mathbf{R}(i,j)$ denote the block matrix in the $i$-th row and $j$-th column, then we have $\mathbf{R}(i, j) = R^{i, i-j}$ with $(i, j)$ indexing from $0$. }

Recall from~\cite{khalil2002nonlinear} that a function $\gamma(\cdot): \mathbb{R}_{\geq 0} \mapsto\mathbb{R}_{\geq 0}$ is a class-$\mathcal{K}$ function if it is continuous, strictly increasing and $\gamma(0) = 0$. The function $\gamma(\cdot)$ is a class-$\mathcal{K}_\infty$ function if it is class-$\mathcal{K}$ and $\lim_{r \rightarrow \infty} \gamma(r) = \infty$. A function $\beta(\cdot, \cdot): \mathbb{R}_{\geq 0} \times \mathbb{R}_{\geq 0} \mapsto \mathbb{R}_{\geq 0}$ is a class-$\mathcal{KL}$ function if for each fixed $t \geq 0$, $\beta(\cdot, t)$ is a class-$\mathcal{K}$ function, and for each fixed $s \geq 0$, $\beta(s, \cdot)$ is decreasing and $\beta(s, t) \rightarrow 0$ as $t \rightarrow \infty$. 


\section{Problem Formulation}
\label{sec:formulation}
Consider an uncertain discrete-time LTI system
\begin{equation}\label{eq:uncertain_system}
\begin{aligned}
x(k+1) = (\hat{A} + \Delta_A) x(k) + (\hat{B} + \Delta_B) u(k) + w(k), \forall k \geq 0
\end{aligned}
\end{equation}
where $x(k) \in \mathbb{R}^{n_x}$ is the system state, $u(k) \in \mathbb{R}^{n_u}$ is the control input, and $w(k) \in \mathbb{R}^{n_x}$ models the additive disturbance at time $k$. The matrices $(\hat{A}, \hat{B})$ denote the nominal dynamics of the system, and $\Delta_A, \Delta_B$ are the parametric model uncertainty. In this work, we assume $\lVert \Delta_A \rVert_\indinf \leq \epsilon_A, \lVert \Delta_B \rVert_\indinf \leq \epsilon_B$ where $\lVert \cdot \rVert_\indinf$ is the $\ell_\infty \rightarrow \ell_\infty$ induced norm, and the additive disturbance $w(k)$ satisfies $\lVert w(k) \rVert_\infty  \leq \sigma_w$ for all $k \geq 0$. For simplicity of notation, we denote these uncertainty sets by 
\begin{equation}
\label{eq:uncertainty_set}
\begin{aligned}
& \mathcal{P}_A = \{\Delta_A \in \mathbb{R}^{n_x \times n_x} \vert \lVert \Delta_A \rVert_\indinf \leq \epsilon_A \}, \quad \mathcal{P}_B = \{\Delta_B \in \mathbb{R}^{n_x \times n_u}\vert \lVert \Delta_B \rVert_\indinf \leq \epsilon_B \},\\ &\mathcal{W} = \{ w \in \mathbb{R}^{n_x} \vert \lVert w \rVert_\infty  \leq \sigma_w \}.
\end{aligned}
\end{equation}

In robust MPC, a finite time constrained robust OCP is solved repeatedly with the current state $x(k)$ as the initial condition, and the first control input in the solution is then applied to the system. Let $T$ denote the horizon of the robust OCP in which we use $x_t, 1 \leq t \leq T$ to denote the predicted states with the initial condition $x_0 = x(k)$. The predicted control inputs $u_t$ and disturbances $w_t$ are defined similarly. 

\begin{problem}[Finite time robust optimal control] \label{prob:robustmpc}
At each time step of the robust MPC, solve the following finite time constrained robust OCP with horizon $T$:
\begin{equation} \label{eq:robustOCP}
\begin{aligned}
\min_{\mathbf{\pi} } & \quad J_T(x(k), \pi)\\
\text{s.t.} &  \quad x_{t + 1} = (\hat{A} + \Delta_A) x_{t} + (\hat{B} + \Delta_B) u_{t} + w_t \\
& \quad u_t = \pi_t(x_{0:t}) \\
& \quad x_t \in \mathcal{X}, u_t \in \mathcal{U}, x_T \in \mathcal{X}_T, t = 0, 1, \cdots, T - 1 \\
& \quad \forall \Delta_A \in \mathcal{P}_A, \forall \Delta_B \in \mathcal{P}_B, \forall w_t \in \mathcal{W}, t = 0, 1, \cdots, T - 1 \\
& \quad  x_0 = x(k)
\end{aligned}
\end{equation}
where the search is done over causal linear time-varying (LTV) state feedback control policies $\mathbf{\pi} = \pi_{0:T-1}$. The sets $\mathcal{X}, \mathcal{U}$, and $\mathcal{X}_T$ are polytopic state, input, and terminal constraints, defined as
\begin{equation*} \label{eq:constraints}
\begin{aligned}
&\mathcal{X} = \lbrace x \in \mathbb{R}^{n} \mid F_x x \leq b_x \rbrace, \  
\mathcal{U} = \lbrace u \in \mathbb{R}^{m} \mid F_u u \leq b_u \rbrace,\\
&\mathcal{X}_T = \lbrace x \in \mathbb{R}^{n} \mid F_T x \leq b_T \rbrace.
\end{aligned}
\end{equation*}
We assume that the sets $\mathcal{X}, \mathcal{U}$, and $\mathcal{X}_T$ are compact and contain the origin in their interior. The definition of the uncertainty sets $\mathcal{P}_A, \mathcal{P}_B,$ and $\mathcal{W}$ are given in~\eqref{eq:uncertainty_set}. The objective function $J_T(x(k), \pi)$ is chosen to be the nominal cost function 
\begin{equation} \label{eq:nominalcost}
\begin{array}{rl}
J_{T}(x(k), \pi) & =  \sum_{t = 0}^{T - 1} (\hat{x}_t^\top Q \hat{x}_t + u_t^\top R u_t ) + \hat{x}_T^\top Q_T \hat{x}_T \\
\text{s.t.} \ & \hat{x}_{t+1} = \hat{A} \hat{x}_t + \hat{B} u_t,  u_t = \pi_t(\hat{x}_{0:t}) \\
& \hat{x}_0 = x(k), \quad \forall t = 0, 1, \cdots, T-1
\end{array}
\end{equation}
with $\hat{x}_{0:T}$ denoting the nominal trajectory. Here $Q \succeq 0, R \succ 0$, and $Q_T \succeq 0$ are the state, input, and terminal weight matrices, respectively.
\end{problem}

In the robust OCP~\eqref{eq:robustOCP}, we aim to find an LTV state feedback controller $\pi_{0:T-1}$ which guarantees the robust satisfaction of all state and input constraints under the norm-bounded model uncertainty and additive disturbances~\eqref{eq:uncertainty_set}. After solving the robust OCP~\eqref{eq:robustOCP}, the first control input $u_0 = \pi_0^*(x_0)$ is applied to drive the system to the next state $x(k+1)$, and the robust OCP~\eqref{eq:robustOCP} is then solved again with $x_0 = x(k+1)$. We denote the MPC control law by $u^k_{MPC}(x(k)) = \pi_0^*(x(k))$. Next, we aim to characterize the closed-loop behavior with $u^k_{MPC}(\cdot)$ in terms of recursive feasibility and closed-loop stability.

\begin{problem}[Closed-loop properties]
\label{prob:cl_properties}
Let $u^k_{MPC}(\cdot)$ denote the robust MPC controller arising from solving the robust OCP~\eqref{eq:robustOCP} at each time step.
\begin{enumerate}
\item Show that $u^k_{MPC}(\cdot)$ is recursively feasible.
\item Show that $u^k_{MPC}(\cdot)$ renders the closed-loop system input-to-state (ISS) stable (the definition of ISS will be given in Section~\ref{sec:ISS}).
\end{enumerate}
\end{problem}

In Section~\ref{sec:SLS} and~\ref{sec:augSLSMPC}, we show how the SLS framework can be used to provide an efficient solution to Problem~\ref{prob:robustmpc}. Then we approach Problem~\ref{prob:cl_properties} by introducing an adaptive horizon strategy for our proposed MPC method. We prove the recursive feasibility of $u^k_{MPC}$ in Section~\ref{sec:recursive} and the ISS of the closed-loop system under $u^k_{MPC}$ in Section~\ref{sec:ISS}. The relevant notations and assumptions for solving each problem are introduced in the corresponding sections.

%

\section{Finite horizon System Level Synthesis}
\label{sec:SLS}
In this section, we introduce relevant concepts from SLS and show how to use finite horizon SLS to transform the robust OCP~\eqref{eq:robustOCP} over linear state feedback controllers into one over closed-loop system responses.

We first consider the LTI system 
\begin{equation}\label{eq:finite_dynamics}
	x_{t + 1} = \hat{A} x_t + \hat{B} u_t + \eta_t
\end{equation} 
over the horizon $t = 0, 1, \cdots, T$ where $\eta_{0:T-1}$ is considered as an additive disturbance. By stacking all the relevant states, control inputs and disturbances as
\begin{equation} \label{eq:concatenation}
\begin{aligned}
&\xx = \begin{bmatrix} x_0^\top & x_1^\top & \cdots & x_T^\top \end{bmatrix}^\top, \quad 
&&\uu = \begin{bmatrix} u_0^\top & u_1^\top & \cdots & u_T^\top \end{bmatrix}^\top, \quad 
&& \bfeta = \begin{bmatrix}
x_0^\top & \eta_0^\top & \cdots  & \eta_{T-1}^\top
\end{bmatrix}^\top,
\end{aligned}
\end{equation}
the system dynamics over the horizon $T$ can be written as
\begin{equation} \label{eq:compact_true_dynamics}
\xx = Z \hat{\sA} \xx + Z \hat{\sB} \uu  + \bfeta
\end{equation}
where $\hat{\sA} = \text{blkdiag}(\hat{A}, \cdots, \hat{A}, 0) \in \mathcal{L}_{TV}^{T, n_x \times n_x}$, $\hat{\sB} = \text{blkdiag}(\hat{B}, \cdots, \hat{B}, 0) \in \mathcal{L}_{TV}^{T, n_x \times n_u}$, and $Z$ is the block-downshift operator with identity matrices of size $n_x \times n_x$ in the first block sub-diagonal and zeros everywhere else. Note that the initial state $x_0$ is embedded as the first component of the disturbance process $\bfeta$. Next, we parameterize an LTV state feedback controller by an operator $\KK \in \mathcal{L}_{TV}^{T, n_u \times n_x}$ with $\uu = \KK \xx$. In other words, at time $t$, we have $u_t = \sum_{i=0}^t K^{t,t-i} x_i$. The closed-loop dynamics of the linear system with the controller $\uu = \KK \xx$ is now given by
\begin{equation} \label{eq:cl_dyn_1}
\xx = Z (\hat{\sA} + \hat{\sB} \KK) \xx  + \bfeta 
\end{equation}
from which we can derive the transfer function from $\bfeta$ to $(\xx, \uu)$ as
\begin{equation}
\label{eq:system_response_K}
\begin{bmatrix}
\xx \\ \uu 
\end{bmatrix} = 
\begin{bmatrix}
(I - Z(\hat{\sA} + \hat{\sB} \KK))^{-1} \\
\KK (I - Z(\hat{\sA} + \hat{\sB} \KK))^{-1}
\end{bmatrix} \bfeta.
\end{equation}
Such maps from $\bfeta$ to $(\xx, \uu)$ are called \emph{system responses}. Since $Z$ is a block-downshift operator, the matrix inversion in~\eqref{eq:system_response_K} is well-defined. Additionally, we observe that the system responses are also block-lower-triangular structure, which allows us to let $\Phix \in \mathcal{L}_{TV}^{T, n_x \times n_x}$ and $\Phiu \in \mathcal{L}_{TV}^{T, n_u \times n_x}$ denote the system responses from $\bfeta$ to $\xx$ and to $\uu$, so that
\begin{equation}\label{eq:system_response_Phi}
\begin{bmatrix}
\xx \\ \uu
\end{bmatrix} = \begin{bmatrix}
\Phix \\ \Phiu
\end{bmatrix} \bfeta.
\end{equation}

In this work, we are particularly interested in the case when $\bfeta$ can be represented as a filtered signal, i.e., $\bfeta = \Sigma \tildeww$ with an invertible matrix $\Sigma \in \mathcal{L}_{TV}^{T, n_x \times n_x}$ and $\tildeww = [x_0^\top \ \widetilde{w}_0^\top  \ \cdots \ \widetilde{w}_{T-1}^\top]^\top$. The map $\Sigma$ can be interpreted as a causal LTV filter acting on the disturbance signal $\tildeww$. A simple example is to choose $\Sigma = \text{blkdiag}(I, \sigma_0 I, \cdots, \sigma_{T-1} I)$ with $\sigma_t  > 0$ and $\lVert \tilde{w}_t \rVert_\infty \leq 1$ for $t = 0, \cdots, T-1$. Then, $\bfeta = \Sigma \tildeww$ models a disturbance signal with time-varying norm bounds $\sigma_t$.
When the filtered signal $\bfeta = \Sigma \tildeww$ is considered, the closed-loop dynamics under the controller $\uu = \KK \xx$ is given by
\begin{equation} \label{eq:dyn_filtered_sig}
\xx = Z (\hat{\sA} + \hat{\sB} \KK) \xx  + \Sigma \widetilde{\ww}
\end{equation}
and the system responses mapping $\tildeww$ to $(\xx, \uu)$ under the controller $\uu = \KK \xx$ are denoted as $\xx = \tildePhix \tildeww$, $\uu =\tildePhiu \tildeww$, respectively. In SLS, the task of controller synthesis is shifted to the design of the closed-loop system responses $\{\tildePhix, \tildePhiu\}$. This is made possible by the following theorem which establishes the relationship between the system responses $\{\tildePhix, \tildePhiu\}$ and the controller $\KK$ that achieves the desired response~\eqref{eq:system_response_Phi}.

\begin{theorem} \label{thm:scaled_equivalence}
	Let $\Sigma \in \mathcal{L}_{TV}^{T, n_x \times n_x}$ be invertible and $\KK \in \mathcal{L}_{TV}^{T, n_u \times n_x}$ be a state feedback controller. Then, for the closed-loop dynamics~\eqref{eq:dyn_filtered_sig} over the horizon $t = 0, \cdots, T$, we have
	\begin{enumerate}
		\item The affine subspace defined by 
		\begin{equation} \label{eq:affine_scaled}
		\begin{bmatrix} I - Z \hat{\sA} & -Z \hat{\sB} \end{bmatrix} 
		\begin{bmatrix} \tildePhix\\ \tildePhiu \end{bmatrix} = \Sigma, \ \tildePhix \in \mathcal{L}_{TV}^{T, n_x \times n_x}, \tildePhiu \in \mathcal{L}_{TV}^{T, n_u \times n_x}
		\end{equation}
		parameterizes all possible system responses $\xx = \tildePhix \widetilde{\ww}, \uu = \tildePhiu \widetilde{\ww}$ for system~\eqref{eq:dyn_filtered_sig}.
		\item For any block-lower-triangular matrices $\{\tildePhix, \tildePhiu \}$ satisfying \eqref{eq:affine_scaled}, the controller $\KK = \tildePhiu \tildePhix^{-1}$ achieves the desired response.
	\end{enumerate}
\end{theorem}

\begin{proof} 1. For a given controller $\KK$, similar to~\eqref{eq:system_response_K}, the system responses from $\tildeww$ to $(\xx, \uu)$ are given by $\tildePhix = (I - Z(\hat{\sA} + \hat{\sB} \KK))^{-1} \Sigma$, $\tildePhiu = \KK (I - Z(\hat{\sA} + \hat{\sB} \KK))^{-1} \Sigma$ and satisfy the affine constraint~\eqref{eq:affine_scaled}.
	
2. First note that the block diagonal of $\tildePhix$ and $\Sigma$ are equal according to~\eqref{eq:affine_scaled}. Then, $\Sigma$ being invertible indicates that $\tildePhix^{-1}$ exists. For any $\{\tildePhix, \tildePhiu\}$ satisfying~\eqref{eq:affine_scaled}, we synthesize the state feedback controller $\KK = \tildePhiu \tildePhix^{-1}$ which satisfies $I - Z\hat{\sA} -Z \hat{\sB} \KK = \Sigma \tildePhix^{-1}$ by multiplying equation~\eqref{eq:affine_scaled} with $\tildePhix^{-1}$ from right. Then we have $(I - Z(\hat{\sA} + \hat{\sB} \KK))^{-1} \Sigma = \tildePhix \Sigma^{-1} \Sigma = \tildePhix$ and $\KK (I - Z(\hat{\sA} + \hat{\sB} \KK))^{-1} \Sigma = \tildePhiu \tildePhix^{-1} \tildePhix = \tildePhiu$.
\end{proof}

\begin{remark}
	\label{remark:affine_constr}
Let $\Sigma = I$ in Theorem~\ref{thm:scaled_equivalence}. Then 
\begin{equation} \label{eq:affine_constr_eta}
	\begin{bmatrix} I - Z \hat{\sA} & -Z \hat{\sB} \end{bmatrix} 
	\begin{bmatrix} \Phix\\ \Phiu \end{bmatrix} = I, \ \Phix \in \mathcal{L}_{TV}^{T, n_x \times n_x}, \Phiu \in \mathcal{L}_{TV}^{T, n_u \times n_x}
\end{equation}
parameterizes all achievable system responses mapping $\bfeta$ to $(\xx, \uu)$, i.e., $\xx = \Phix \bfeta, \uu = \Phiu \bfeta$. This recovers the result in~\cite[Theorem 2.1]{ANDERSON2019364} which does not consider $\bfeta$ as a filtered signal. By plugging in $\bfeta = \Sigma \tildeww$, we have $\xx = \Phix \bfeta = \Phix \Sigma \tildeww = \tildePhix \tildeww$ and $\uu = \Phiu \bfeta = \Phiu \Sigma \tildeww = \tildePhiu \tildeww$. Therefore, we obtain the mapping between the system responses with respect to $\bfeta$ and $\tildeww$ as  
\begin{equation}
	\tildePhix = \Phix \Sigma, \quad \tildePhiu = \Phiu \Sigma.
\end{equation}
In addition, both system responses $\{\Phix, \Phiu \}$ and $\{\tildePhix, \tildePhiu \}$ are achievable by the same state feedback controller $\KK = \tildePhiu \tildePhix^{-1} = (\Phiu \Sigma) (\Phix \Sigma)^{-1} = \Phiu \Phix^{-1}$.


\end{remark}


Theorem~\ref{thm:scaled_equivalence} states any LTV state feedback controller $\KK$ can be parameterized by a pair of system responses $\{\tildePhix, \tildePhiu\}$ satisfying the affine constraint~\eqref{eq:affine_scaled}. Therefore, the search for the controller $\KK$ can be transformed into the search for the closed-loop system responses $\{\tildePhix, \tildePhiu\}$ with an additional affine constraint~\eqref{eq:affine_scaled}. The maps $\xx = \tildePhix \tildeww, \uu = \tildePhiu \tildeww$ transparently describe the effects of the disturbance $\tildeww$ on $\xx$ and $\uu$ under a state feedback controller; this can be used to translate any constraint on $(\xx, \uu)$ into an equivalent one on $\{\tildePhix, \tildePhiu\}$. See~\cite{ANDERSON2019364} for examples of solving control problems through SLS. Note that the constraint~\eqref{eq:affine_scaled} is jointly affine in $(\tildePhix, \tildePhiu, \Sigma)$. By representing $\bfeta = \Sigma \tildeww$, we can use the filter $\Sigma$ as a design parameter in addition to $\{\tildePhix, \tildePhiu \}$ in robust controller synthesis. 

In the next section, we take the model uncertainty $(\Delta_A, \Delta_B)$ from~\eqref{eq:uncertain_system} into account and propose an SLS-based convex inner approximation of the robust OCP~\eqref{eq:robustOCP} which jointly optimizes over LTV state feedback controllers and upper bounds on the deviation of the system states from the nominal predicted trajectories through the parameterization of the system responses $\{\tildePhix, \tildePhiu\}$ and the filter $\Sigma$.

%

\section{Robust MPC through convex inner approximation}
\label{sec:augSLSMPC}
Inspired by~\cite{bujarbaruah2021simple}, we derive a convex inner approximation of the robust OCP~\eqref{eq:robustOCP} in the space of system responses by describing the uncertain system dynamics as
\begin{equation}
\label{eq:aug_dist_dyn}
\begin{aligned}
x_{t + 1} & = \hat{A} x_{t}  + \hat{B} u_{t} + \Delta_A x_{t}  + \Delta_B u_{t} + w_t\\
&= \hat{A} x_{t}  + \hat{B} u_{t} + \eta_t,
\end{aligned}
\end{equation}
where $\eta_t := \Delta_A x_{t}  + \Delta_B u_{t} + w_t$ lumps the effects of the model uncertainty $\{\Delta_A, \Delta_B\}$ and additive disturbance $w_t$. The perturbation $\eta_t$ can be interpreted as the deviation of the system from the predicted nominal trajectory. We call the signal $\eta_{0:T-1}$ the \emph{\augdist{}} to distinguish it from the additive disturbance signal $w_{0:T-1}$. For robust control, we want to bound the lumped uncertainty $\eta_{0:T-1}$ in the presence of the feedback controller to be designed. Since each $\eta_t$ is a function of the states $x_{0:t}$, the disturbances $w_{0:t}$ and the underlying controller $\KK$, the bounds on $\eta_t$ must be time-varying in order to be tight. In our proposed method, at each time step we solve a robust OCP that jointly optimizes over the state feedback controller $\KK$ and the time-varying upper bounds on $\eta_{0:T-1}$. The derivation of our proposed robust OCP is presented step-by-step below.


\subsection{Characterization of the \augdist{}}
\label{sec:aug_dist}
Recall that $\hat{\sA} = \text{blkdiag}(\hat{A}, \cdots, \hat{A}, 0) $, $\hat{\sB} = \text{blkdiag}(\hat{B}, \cdots, \hat{B}, 0)$. We concatenate the \augdists{} $\eta_t$ over the horizon $T$ as shown in~\eqref{eq:concatenation}. By Theorem~\ref{thm:scaled_equivalence}, the affine subspace
\begin{equation} \label{eq:nominal_affine}
\begin{bmatrix} I - Z \hat{\sA} & -Z \hat{\sB} \end{bmatrix} 
\begin{bmatrix} \Phix \\ \Phiu \end{bmatrix} = I, \ \Phix \in \mathcal{L}_{TV}^{T, n_x \times n_x}, \Phiu \in \mathcal{L}_{TV}^{T, n_u \times n_x}
\end{equation}
parameterizes all achievable closed-loop system responses that map the \augdist{} $\bfeta$ to $\xx$ and $\uu$ as shown in Remark~\ref{remark:affine_constr}. By stacking the model uncertainty parameters over the horizon $T$ as $\DDelta_A = \text{blkdiag}(\Delta_A, \cdots, \Delta_A, 0) \in \mathcal{L}_{TV}^{T, n_x \times n_x}$, $\DDelta_B = \text{blkdiag}(\Delta_B, \cdots, \Delta_B, 0) \in \mathcal{L}_{TV}^{T, n_x \times n_u}$, we have that $\bfeta = Z \DDelta_A \xx + Z \DDelta_u \uu + \ww$ from the definition of \augdist{} in~\eqref{eq:aug_dist_dyn}. Combined with the achievable system responses $\{\Phix, \Phiu\}$ given in~\eqref{eq:nominal_affine}, we describe the dynamics of $\bfeta$ as 
\begin{equation} \label{eq:dyn_aug}
\begin{aligned}
\bfeta &= Z \begin{bmatrix}
\DDelta_A & \DDelta_B
\end{bmatrix} 
\begin{bmatrix}
\xx \\ \uu
\end{bmatrix} + \ww = Z \begin{bmatrix}
\DDelta_A & \DDelta_B
\end{bmatrix} 
\begin{bmatrix}
\Phix \\ \Phiu
\end{bmatrix} \bfeta + \ww.
\end{aligned}
\end{equation}
We can view equations~\eqref{eq:system_response_Phi} and~\eqref{eq:dyn_aug} as a linear dynamical system with $\eta_t$ as the state, $w_t$ as the exogenous input and $x_t, u_t$ as the system output. Note that $Z[\DDelta_A \ \DDelta_B] [\Phix^\top \ \Phiu^\top]^\top$ is strictly causal, and therefore the dynamics of $\bfeta$ in~\eqref{eq:dyn_aug} is well-defined. In the next subsection, we propose an inner-approximation of the robust OCP~\eqref{eq:robustOCP} based on the dynamics of the \augdist{} $\bfeta$.

\begin{remark}
From~\eqref{eq:system_response_Phi} and~\eqref{eq:dyn_aug} we obtain the map $\ww \mapsto (\xx, \uu)$:
\begin{equation} \label{eq:response_true}
	\begin{bmatrix}
\xx \\ \uu
\end{bmatrix} = \begin{bmatrix}
\Phix \\ \Phiu
\end{bmatrix} (I - Z \begin{bmatrix}
\DDelta_A & \DDelta_B
\end{bmatrix} 
\begin{bmatrix}
\Phix \\ \Phiu
\end{bmatrix} )^{-1} \ww.
\end{equation}
Eqn.~\eqref{eq:response_true} describes the system responses from $\ww$ to $(\xx, \uu)$ under the model uncertainty $(\Delta_A, \Delta_B)$. Such a relationship has been exploited in~\cite{dean2019safely, chen2020robust} to solve robust OCPs through SLS. However, the complex structure of the system responses in~\eqref{eq:response_true} makes it difficult to assess the effects of the system uncertainty and thus renders the resultant robust MPC method conservative. We provide a detailed comparison between our method and~\cite{chen2020robust} in Section~\ref{sec:simulation}.
\end{remark}

\subsection{Inner approximation of the robust OCP}
\label{sec:tightening}
In this subsection, we propose a convex inner approximation of the robust OCP~\eqref{eq:robustOCP} by jointly optimizing over the closed-loop system responses and the norm bounds on the \augdist{} $\bfeta$. Our proposed method is motivated by the following observations:
\begin{enumerate}
\item By over-approximating the \augdists{} $\eta_{0:T-1}$ by $\ell_\infty$ norm balls, we can easily tighten the polytopic state and control input constraints of the robust OCP~\eqref{eq:robustOCP} through H\"{o}lder's inequality as in the additive disturbance case~\cite{goulart2006optimization, dean2019safely, bujarbaruah2021simple}.
\item To reduce the conservatism of the $\ell_\infty$ ball over-approximation, it is desirable to upper bound $\lVert \eta_t \rVert_\infty$ for each time step $t$ separately based on the interdependence between the \augdist{} $\bfeta$ and $(\xx, \uu)$.
\end{enumerate}

Next, we first over-approximate the \augdist{} $\bfeta$ using $\ell_\infty$ norm balls, and then derive norm upper bounds of these $\ell_\infty$ balls by exploiting the dynamics of $\bfeta$. Finally, we obtain a convex inner approximation of the robust OCP~\eqref{eq:robustOCP} by tightening the state and input constraints based on the $\ell_\infty$ ball over-approximation of $\bfeta$.

\subsubsection{Over-approximating \augdist{} through $\ell_\infty$ balls}
We first assume each \augdist{} $\eta_t$ can be over-approximated by an $\ell_\infty$ norm ball with radius $\sigma_t$, i.e., $\lVert \eta_t \rVert_\infty \leq \sigma_t$ for $0 \leq t \leq T-1$. Equivalently, we can represent $\eta_t$ as $\eta_t  = \sigma_t \widetilde{w}_t$ with $\lVert \widetilde{w}_t \rVert_\infty \leq 1$. Let $\tildeww = [x_0^\top \ \widetilde{w}_0^\top \ \cdots \ \widetilde{w}_{T-1}^\top]^\top$ and $\Sigma = \text{blkdiag}(I, \sigma_0 I, \cdots, \sigma_{T-1} I)$. Then the \augdist{} $\bfeta$ can be considered as the filtered signal $\bfeta = \Sigma \tildeww$. By Theorem~\ref{thm:scaled_equivalence}, we have  
\begin{equation} \label{eq:scaled_affine}
\begin{bmatrix} I - Z \hat{\sA} & -Z \hat{\sB} \end{bmatrix} 
\begin{bmatrix} \tildePhix \\ \tildePhiu \end{bmatrix} = \Sigma, \ \tildePhix \in \mathcal{L}_{TV}^{T, n_x \times n_x}, \tildePhiu \in \mathcal{L}_{TV}^{T, n_u \times n_x}
\end{equation}
parameterizes all achievable closed-loop system responses $\{\tildePhix, \tildePhiu \}$ that map $\tildeww$ to $(\xx, \uu)$, i.e., $\xx = \tildePhix \tildeww, \uu = \tildePhiu \tildeww$. As a result, the dynamics~\eqref{eq:dyn_aug} of the \augdist{} $\bfeta$ can be equivalently written as
\begin{equation}\label{eq:dyn_aug_norm}
\bfeta = Z \begin{bmatrix}
\DDelta_A & \DDelta_B
\end{bmatrix} 
\begin{bmatrix}
\tildePhix \\ \tildePhiu
\end{bmatrix} \tildeww + \ww.
\end{equation}
Compared with~\eqref{eq:dyn_aug}, the representation~\eqref{eq:dyn_aug_norm} reveals the dependence of the \augdist{} $\bfeta$ on the normalized disturbance signal $\tildeww$ where the scale of each $\eta_t$ is absorbed in $\{\tildePhix, \tildePhiu \}$ by constraint~\eqref{eq:scaled_affine} and the construction of $\Sigma$. Since constraint~\eqref{eq:scaled_affine} is affine in $\{\tildePhix, \tildePhiu, \Sigma \}$, we can jointly optimize over the closed-loop system responses $\{\tildePhix, \tildePhiu \}$ and the norm bounds $\{\sigma_t \}_{t=0}^{T-1}$ encoded in $\Sigma$ while maintaining the convexity of the overall formulation.

\subsubsection{Upper-bounding the magnitude of lumped uncertainty}
Recall that we let $\{\sigma_0,\sigma_1, \cdots, \sigma_{T-1}\}$ be a set of upper bounds on the $\ell_\infty$ norm of the \augdists{} $\eta_t$, i.e., $\lVert \eta_t \rVert_\infty \leq \sigma_t$ for $t = 0, \cdots, T-1$. From Eqn.~\eqref{eq:dyn_aug_norm}, we have $\eta_0 = \Delta_A \widetilde{\Phi}_x^{0,0} x_0 + \Delta_B \widetilde{\Phi}_u^{0,0} x_0 + w_0$, and for $1 \leq t \leq T-1$, 
\begin{equation}
\eta_{t} = \Delta_A (\widetilde{\Phi}_x^{t,t}x_0 + \sum_{i=1}^{t} \widetilde{\Phi}_x^{t,t-i} \widetilde{w}_{i-1}) + \Delta_B(\widetilde{\Phi}_u^{t,t} x_0 + \sum_{i=1}^{t} \widetilde{\Phi}^{t,t-i} \widetilde{w}_{i-1}) + w_t.
\end{equation}
By the triangle inequality and submultiplicativity of $\lVert \cdot \rVert_\infty$, we can upper bound $\lVert \eta_{t} \rVert_\infty$ by
\begin{equation} \label{eq:aug_dist_ub}
\begin{aligned}
\lVert \eta_{t} \rVert_\infty & \leq \lVert \Delta_A \rVert_\indinf (\lVert \widetilde{\Phi}_x^{t,t}x_0 \rVert_\infty  + \sum_{i=1}^t \lVert \widetilde{\Phi}_x^{t,t-i} \rVert_\indinf \lVert \widetilde{w}_{i-1} \rVert_\infty) + \\
& \quad \quad \quad \quad \lVert \Delta_B \rVert_\indinf (\lVert \widetilde{\Phi}_u^{t,t}x_0 \rVert_\infty  + \sum_{i=1}^t \lVert \widetilde{\Phi}_u^{t,t-i} \rVert_\indinf \lVert \widetilde{w}_{i-1} \rVert_\infty) + \lVert w_t \rVert_\infty  \\
& \leq \epsilon_A (\lVert \widetilde{\Phi}_x^{t,t}x_0 \rVert_\infty  + \sum_{i=1}^t \lVert \widetilde{\Phi}_x^{t,t-i} \rVert_\indinf) + \epsilon_B (\lVert \tilde{\Phi}_u^{t,t}x_0 \rVert_\infty  + \sum_{i=1}^t \lVert \widetilde{\Phi}_u^{t,t-i} \rVert_\indinf) + \sigma_w
\end{aligned}
\end{equation}
using $\lVert \widetilde{w}_t \rVert_\infty \leq 1$ for $0 \leq t \leq T-1$. The upper bound~\eqref{eq:aug_dist_ub} on the norm of the \augdist{} at time $t$ depends on the system uncertainty parameters $\epsilon_A, \epsilon_B, \sigma_w$, and the closed-loop system responses up to time $t-1$. For $t=0$, the upper bound on $\lVert \eta_0 \rVert_\infty$ can be derived similarly.

\begin{lemma} \label{lem:suff_cond}
For the closed-loop system responses $\{\tildePhix, \tildePhiu \}$ satisfying constraint~\eqref{eq:scaled_affine} with a set of positive real numbers $\{\sigma_t\}_{t=0}^{T-1}$, and the uncertainty parameters described in~\eqref{eq:uncertainty_set}, a sufficient condition for $\{\sigma_t\}_{t=0}^{T-1}$ to be upper bounds on the norm of the \augdist{} $\eta_t$ is given by 
\begin{equation} \label{eq:suff_cond_ub}
\begin{aligned}
&\epsilon_A \lVert \widetilde{\Phi}_x^{0,0} x_0 \rVert_\infty + \epsilon_B \lVert \widetilde{\Phi}_u^{0,0} x_0 \rVert_\infty + \sigma_w \leq \sigma_0 \\
&\epsilon_A (\lVert \widetilde{\Phi}_x^{t,t}x_0 \rVert_\infty  + \sum_{i=1}^t \lVert \widetilde{\Phi}_x^{t,t-i} \rVert_\indinf) + \epsilon_B (\lVert \widetilde{\Phi}_u^{t,t}x_0 \rVert_\infty  + \sum_{i=1}^t \lVert \widetilde{\Phi}_u^{t,t-i} \rVert_\indinf) + \sigma_w \leq \sigma_{t}
\end{aligned}
\end{equation}
for $t = 1, \cdots, T-1$. In other words, constraints~\eqref{eq:suff_cond_ub} guarantee that $\lVert \eta_t \rVert_\infty \leq \sigma_t$ for $0 \leq t \leq T-1 $.
\end{lemma}
\begin{proof}
Since $\eta_0 = \Delta_A \widetilde{\Phi}_x^{0,0} x_0 + \Delta_B \widetilde{\Phi}_u^{0,0} x_0 + w_0$, by the triangle inequality and the submultiplicativity of the $\ell_\infty$ norm, the first inequality in~\eqref{eq:suff_cond_ub} guarantees $\lVert \eta_0 \rVert_\infty \leq \sigma_0$. Then, by induction and the upper bounds derived in~\eqref{eq:aug_dist_ub}, $\lVert \eta_t \rVert_\infty \leq \sigma_t$ holds for $t =1, \cdots, T-1$.
\end{proof}

\subsubsection{Constraint tightening of the robust OCP}
With the system responses $\{\tildePhix, \tildePhiu \}$ characterized by constraint~\eqref{eq:scaled_affine}, the map from the normalized disturbances $\tildeww$ to $(\xx, \uu)$ is given by $\xx = \tildePhix \tildeww$ and $\uu = \tildePhiu \tildeww$, which allows us to tighten the state and control input constraints of the robust OCP~\eqref{eq:robustOCP}.
We illustrate the constraint tightening procedure using the state constraint $x_t \in \mathcal{X}$ -- an analogous argument can be applied to the control and terminal constraint sets. Recall that $\mathcal{X} = \{x \vert F_x x \leq b_x\}$ and $F_x \in \mathbb{R}^{n_\mathcal{X} \times n_x}$, i.e., there are $n_\mathcal{X}$ linear constraints in defining the set $\mathcal{X}$. We use $f^\top x \leq b$ to denote an arbitrary linear constraint in the definition of $\mathcal{X}$ and introduce the notation $(f, b) \in \text{facet}( \mathcal{X})$ with $\text{facet}(\mathcal{X}) = \{(F_x(i,:), b(i)) \vert i = 1, \cdots, n_\mathcal{X}\}$ being the set of all linear constraint parameters of $\mathcal{X}$. Since $x_t = \widetilde{\Phi}_x^{t,t} x_0 + \sum_{i = 1}^{t} \widetilde{\Phi}_x^{t,t-i} \widetilde{w}_{i-1}$, the tightening of the constraint $x_t \in \mathcal{X}$ is given by 
\begin{equation} \label{eq:state_tightening}
f^\top \widetilde{\Phi}_x^{t,t} x_0 + \sum_{i=1}^t \lVert f^\top \widetilde{\Phi}_x^{t,t-i} \rVert_1 \leq b, \quad \forall (f, b) \in \text{facet}(\mathcal{X}), \quad t = 0, \cdots, T-1.
 \end{equation}
This is due to $f^\top \widetilde{\Phi}_x^{t,t-i} \widetilde{w}_{i-1} \leq \lVert f^\top \widetilde{\Phi}_x^{t,t-i} \rVert_1 \lVert \widetilde{w}_{i-1}\rVert_\infty \leq \lVert f^\top \widetilde{\Phi}_x^{t,t-i} \rVert_1$ where the first inequality follows from H\"older's inequality and the second is from the fact $\lVert \widetilde{w}_t \rVert_\infty \leq 1$ for $0 \leq t \leq T-1$. Similarly, the tightened constraints for the terminal state and control inputs are given by 
\begin{align}
& f^\top \widetilde{\Phi}_x^{T,T} x_0 + \sum_{i=1}^T \lVert f^\top \widetilde{\Phi}_u^{T,T-i} \rVert_1 \leq b, \quad \forall (f, b) \in \text{facet}(\mathcal{X}_T), \label{eq:terminal_tightening}\\
& f^\top \widetilde{\Phi}_u^{t,t} x_0 + \sum_{i=1}^t \lVert f^\top \widetilde{\Phi}_u^{t,t-i} \rVert_1  \leq b, \quad \forall (f, b) \in \text{facet}(\mathcal{U}), \quad t = 0, \cdots, T-1. \label{eq:input_tightening}
\end{align}

\begin{theorem}
Consider the following convex quadratic program
\begin{equation} \label{eq:tightening}
\begin{aligned}
\underset{\tildePhix, \tildePhiu, \{\sigma_t\}_{t=0}^{T-1}}{\text{minimize}} & \quad \Big \lVert \begin{bmatrix}
\sQ^{1/2} & \\ & \sR^{1/2}
\end{bmatrix} \begin{bmatrix}
\tildePhix(:,0) \\ \tildePhiu(:,0)
\end{bmatrix}x_0 \Big \rVert_2^2 \\
\text{subject to} & \quad \begin{bmatrix} I - Z \hat{\sA} & -Z \hat{\sB} \end{bmatrix} 
\begin{bmatrix} \tildePhix \\ \tildePhiu \end{bmatrix} = \Sigma, \ \tildePhix \in \mathcal{L}_{TV}^{T, n_x \times n_x}, \tildePhiu \in \mathcal{L}_{TV}^{T, n_u \times n_x}, \\
& \quad \Sigma = \text{blkdiag}(I, \sigma_0 I, \cdots, \sigma_{T-1} I), \\
& \quad \text{disturbance bound constraints } \eqref{eq:suff_cond_ub}, \\
& \quad \text{tightened constraints } \eqref{eq:state_tightening}, \eqref{eq:terminal_tightening}, \text{and } \eqref{eq:input_tightening},\\
& \quad x_0 = x(k),
\end{aligned}
\end{equation}
where $\sQ= \text{blkdiag}(Q, \cdots, Q, Q_T)$, and $\sR = \text{blkdiag}(R, \cdots, R, 0)$ encode the weights of the stage and terminal costs in Eqn.~\eqref{eq:nominalcost}, respectively. For any feasible solution $\{\tildePhix, \tildePhiu, \{\sigma_t\}_{t=0}^{T-1} \}$ of problem~\eqref{eq:tightening}, the synthesized controller $\KK = \tildePhiu \tildePhix^{-1}$ guarantees the robust satisfaction of all state and control input constraints of the robust OCP~\eqref{eq:robustOCP}. 
\end{theorem}
\begin{proof}
The proof follows from Theorem~\ref{thm:scaled_equivalence} and the derivation of the constraints~\eqref{eq:suff_cond_ub} to~\eqref{eq:input_tightening} presented in this subsection. First, from constraint~\eqref{eq:suff_cond_ub}, we have $\{\sigma_t\}_{t=0}^{T-1}$ are all lower-bounded by the norm of the additive disturbance $\sigma_w >0$. Therefore, any solution $\Sigma$ of problem \eqref{eq:tightening} is invertible and Theorem~\ref{thm:scaled_equivalence} can be applied. By Theorem~\ref{thm:scaled_equivalence} and Lemma~\ref{lem:suff_cond}, the affine constraint~\eqref{eq:scaled_affine} and the upper bound constraint~\eqref{eq:suff_cond_ub} guarantee that $\lVert \eta_t \rVert_\infty \leq \sigma_t$ under the synthesized controller $\KK =  \tildePhiu \tildePhix^{-1}$. Then, constraints \eqref{eq:state_tightening} to \eqref{eq:input_tightening} guarantee robust constraint satisfaction of the robust OCP~\eqref{eq:robustOCP} by tightening the state and control input constraints based on the $\ell_\infty$ norm ball over-approximation of the \augdist{} $\bfeta$. By letting $\eta_t = 0$ for $0 \leq t \leq T-1$, we have $\Sigma = \text{blkdiag}(I, 0, \cdots, 0)$ and $\tildePhix(:,0) = \Phix(:,0), \tildePhiu(:,0) = \Phiu(:,0)$ by Remark~\ref{remark:affine_constr}. Therefore, $\tildePhix(:,0) x_0 = \Phix(:,0) x_0$ and $\tildePhiu(:,0)x_0 = \Phiu(:,0) x_0$ denote the nominal states and control inputs, respectively, and the objective function in~\eqref{eq:tightening} coincides with the nominal cost function given in~\eqref{eq:nominalcost}.
\end{proof}

Problem~\eqref{eq:tightening} is jointly convex over the closed-loop system responses $\{\tildePhix, \tildePhiu\}$ and the norm upper bounds on the \augdist{} $\{\sigma_t\}_{t=0}^{T-1}$, allowing them to be simultaneously optimized. We call problem~\eqref{eq:tightening} a convex inner approximation of the robust OCP~\eqref{eq:robustOCP} since the constraints in~\eqref{eq:tightening} define an inner approximation to the feasible set of the robust OCP~\eqref{eq:robustOCP}. Therefore, problem~\eqref{eq:tightening} is in general more conservative than the robust OCP \eqref{eq:robustOCP}. However, we note that in the special case of horizon $T = 1$, problem~\eqref{eq:tightening} is non-conservative in that problem~\eqref{eq:tightening} solves the robust OCP~\eqref{eq:robustOCP} exactly. 

\begin{lemma} \label{lem:tightness}
For horizon $T=1$, the optimization problem \eqref{eq:tightening} is a non-conservative inner approximation of the robust OCP~\eqref{eq:robustOCP} over LTV state feedback controllers. 
\end{lemma}
\begin{proof}
The proof relies on showing that any feasible solution of the robust OCP~\eqref{eq:robustOCP} over state feedback controllers constructs a feasible solution of the inner approximation~\eqref{eq:tightening} when $T=1$. This is because when $T=1$, the upper bound $\sigma_0$ on $\lVert \eta_0 \rVert_\infty$ given by~\eqref{eq:aug_dist_ub} is tight. For details of the proof, see Appendix~\ref{proof:tightness}.
\end{proof}

For horizons $T > 1$, the convex inner approximation~\eqref{eq:tightening} is conservative because in general there are not enough degrees of freedom to choose  $\Delta_A, \Delta_B$ and $\widetilde{w}_{0:t-1}$ adversarially to achieve the upper bounds on $\lVert \bar{w}_t \rVert_\infty$ in~\eqref{eq:aug_dist_ub}. We note that for a feasible solution of problem~\eqref{eq:tightening}, under the synthesized robust controller $\KK= \tildePhiu \tildePhix^{-1}$, all possible trajectories of the uncertain system are contained in a tube consisting of $\ell_\infty$ balls centered around the nominal trajectory with varying radii $\{\sigma_t \}_{t=0}^{T-1}$. The $\ell_\infty$ ball is a natural choice in bounding the \augdist{} under the uncertainty parameter assumption~\eqref{eq:uncertainty_set} and it allows efficient constraint tightening in problem~\eqref{eq:tightening}. In comparison, the tube-MPC method~\cite{langson2004robust} does not consider LTV state feedback controllers; instead, it parameterizes feedback controllers implicitly by a tube with fixed shape, and uses vertex enumeration over both the uncertainty parameter and the tube to tighten the constraints.

Finally, our proposed MPC controller is given by
\begin{equation}
\label{eq:mpc_policy}
u^k_{MPC}(x(k)) = \widetilde{\Phi}_u^{0,0^*} x(k)
\end{equation}
with $\widetilde{\Phi}_u^{0,0^*}$ being the $\widetilde{\Phi}_u^{0,0}$-component of the solution of~\eqref{eq:tightening}. Since the robust optimal control problem~\eqref{eq:tightening} is obtained by tightening the constraints using the \augdist{}, we call our proposed method \emph{\augdist{} SLS MPC}. 

\begin{remark}
When $\epsilon_A = \epsilon_B = 0$, the formulation~\eqref{eq:tightening} recovers the SLS-based robust MPC method shown in~\cite{sieber2021system, dean2019safely}, as the upper bound constraint~\eqref{eq:suff_cond_ub} is trivially satisfied. In this special case where only additive disturbances are present, the \augdist{} SLS MPC is equivalent to the affine state feedback\footnote{An affine state-feedback controller is used in~\cite{goulart2006optimization} while SLS only parameterizes a linear state feedback controller. However, we can always augment system~\eqref{eq:uncertain_system} with state $\bar{x} = [x; 1]$ to make them equivalent.} or disturbance feedback MPC approaches~\cite{goulart2006optimization}, as shown in~\cite{sieber2021system}. 
\end{remark}

\section{Recursive feasibility}
\label{sec:recursive}
Our robust MPC policy~\eqref{eq:mpc_policy} is obtained by solving the convex inner approximation~\eqref{eq:tightening} of the robust OCP~\eqref{eq:robustOCP} at each time $k$. Although our goal is to find a robust state feedback controller $\KK$ through~\eqref{eq:tightening}, the optimization is over the system responses with tightened constraints. As a result, even if a controller $\KK$ is a feasible solution to the robust OCP~\eqref{eq:robustOCP}, there is no guarantee that the system responses $\{\tildePhix, \tildePhiu \}$ it generates is feasible for the convex inner approximation~\eqref{eq:tightening}. This prevents us from applying standard controller shifting argument~\cite[Chapter 12]{borrelli2017predictive} to prove the recursive feasibility of our proposed robust MPC method when the horizon $T$ is constant. 



To overcome this difficulty, we allow the horizon of the robust MPC problem~\eqref{eq:robustOCP} to vary with time. We denote $T_k$ as the MPC horizon at time $k$. To guarantee the recursive feasibility of \augdist{} SLS MPC, we use an adaptive horizon strategy described as  
\begin{equation} \label{eq:adaptive_horizon}
T_k = \argmin_{T \in \{1, \cdots, T_{max}\}} J_T^*(x(k))
\end{equation}
where $T_{max}$ is a fixed horizon upper bound and $J_T^*(x(k))$ is the optimal objective of~\eqref{eq:tightening} with horizon $T$. We let $J^*_T(x(k)) = \infty$ if the convex tightening~\eqref{eq:tightening} is infeasible. The following assumptions about the terminal set are made to ensure recursive feasibility.

\begin{assumption} \label{assump:robust_stable}
	A linear state feedback controller $\kappa(x) = Kx$ is known such that the $\hat{A} + \Delta_A + (\hat{B} + \Delta_B) K$ is stable for all $\Delta_A \in \mathcal{P}_A, \Delta_B \in \mathcal{P}_B$.
\end{assumption}

\begin{assumption} \label{assump:robust_invariant}
	The terminal set $\mathcal{X}_T$ is a robust forward invariant set under the local controller $\kappa(x) = Kx$, i.e., for all $x \in \mathcal{X}_T$, we have $x \in \mathcal{X}, Kx \in \mathcal{U}$, and $((\hat{A} + \Delta_A) + (\hat{B} + \Delta_B) K)x + w \in \mathcal{X}_T$ for all $\Delta_A \in \mathcal{P}_A, \Delta_B \in \mathcal{P}_B, w \in \mathcal{W}$.
\end{assumption}

These two assumptions are standard in the analysis of MPC and computational tools~\cite[Chapter 10]{borrelli2017predictive} are available to compute the local feedback controller $\kappa(x)=Kx$ and find the required terminal set $\mathcal{X}_T$. We first show that, when \augdist{} SLS MPC is feasible at $x(0)$ with horizon $T_0 \geq 2$, solving the convex inner approximation~\eqref{eq:tightening} in a decreasing horizon manner drives the system into the terminal set in $T_0$ steps. 

\begin{theorem} \label{thm:recursive_feasibility}
Let the robust OCP inner approximation~\eqref{eq:tightening} be feasible at time $k$ with horizon $T_k \in \{2, 3, \cdots,\allowdisplaybreaks T_{max}\}$ and $x_0 = x(k)$. Let $x(k+1)$ be the state at time $k+1$ with the MPC controller~\eqref{eq:mpc_policy}. Then the inner approximation~\eqref{eq:tightening} is feasible at time $k+1$ with horizon $T_{k+1} = T_k -1$ and $x_0 = x(k+1)$.
\end{theorem}

\begin{proof}
The proof relies on an explicit construction of a feasible solution $\tildePhix, \tildePhiu, \{\sigma_t\}_{t=0}^{T_{k+1}-1}$ for problem~\eqref{eq:tightening} with the initial condition $x(k+1)$ and horizon $T_{k+1}$. See Appendix \ref{proof:feasibility} for details.
\end{proof}


By Theorem~\ref{thm:recursive_feasibility}, problem~\eqref{eq:tightening} is always feasible with $T_k = T_0 - k \geq 0$ at $x(k)$. As a result, at time $k =T_0 -1$, the \augdist{} SLS MPC with horizon $T_k = 1$ is feasible and guarantees the next state $x(T_0)$ will lie in the terminal set $\mathcal{X}_T$. Now with the construction of the terminal set, we can guarantee that $u_{MPC}^k(\cdot)$ is recursively feasible inside the terminal set with horizon $T_k = 1$ and renders the terminal set robustly forward invariant.

\begin{lemma} \label{lem:terminal_invariance}
Under Assumption~\ref{assump:robust_stable} and~\ref{assump:robust_invariant}, for any state $x(k) \in \mathcal{X}_T$, the convex inner approximation~\eqref{eq:tightening} is feasible with horizon $T_k = 1$ and $x(k)$ as the initial condition. In addition, the terminal set $\mathcal{X}_T$ is a robust forward invariant set under the \augdist{} SLS MPC controller $u_{MPC}^k(\cdot)$ with horizon $T_k = 1$.
\end{lemma}
\begin{proof}
The recursive feasibility of $u_{MPC}^k(\cdot)$ inside the terminal set holds because the inner approximation~\eqref{eq:tightening} with horizon one is not conservative (see Lemma~\ref{lem:tightness}) and a terminal set $\mathcal{X}_T$ and local controller $\kappa(x)$ exist which satisfy Assumptions~\ref{assump:robust_stable} and~\ref{assump:robust_invariant}. With horizon $T_k = 1$, \augdist{} SLS MPC guarantees the next state $x(k+1)$ is always in $\mathcal{X}_T$, proving the robust forward invariance of the terminal set $\mathcal{X}_T$.
\end{proof}

\begin{theorem}[Recursive feasibility]
A \augdist{} SLS MPC policy which solves the robust OCP inner approximation~\eqref{eq:tightening} and adopts the adaptive horizon strategy~\eqref{eq:adaptive_horizon} is recursively feasible. In other words, if problem~\eqref{eq:tightening} is feasible at time $k=0$ with horizon $T_0$, it is feasible all time steps $k \geq 1$ with horizon $T_k$ selected by~\eqref{eq:adaptive_horizon}.
\end{theorem}
\begin{proof}
We divide the discussion into two cases. At time $k$, if the inner approximation~\eqref{eq:tightening} is feasible with horizon $T_k \geq 2$, then by Theorem~\ref{thm:recursive_feasibility} it is feasible at time $k+1$ with horizon $T_{k+1} = T_k -1$. We refer the reader to the proof of Theorem~\ref{thm:recursive_feasibility} for the construction of a feasible solution of problem~\eqref{eq:tightening} at time $k+1$. If at time $k$, problem~\eqref{eq:tightening} is feasible with horizon $T_k = 1$, then by the proof of Lemma~\ref{lem:terminal_invariance} we know that the local controller $\kappa(x) = Kx$ from Assumption~\ref{assump:robust_invariant} yields a feasible solution to \eqref{eq:tightening} at time $k+1$ with horizon $T_{k+1} = 1$. Summarizing the above two cases, we conclude that \augdist{} SLS MPC is recursively feasible.
\end{proof}

\subsection{Horizon design choices}
Adaptive horizon strategies have been applied to guarantee recursive feasibility in various MPC design methods such as tube MPC~\cite{langson2004robust}, learning MPC~\cite{rosolia2021robust}, and tightening-based MPC~\cite{bujarbaruah2020robust}. In addition to the strategy~\eqref{eq:adaptive_horizon} which chooses the horizon that minimizes the cost function along the prediction horizon, other horizon selection procedures are available, such as the decreasing horizon~\cite{langson2004robust}, minimum time~\cite{krener2018adaptive, goulart2006optimization}, and single policy~\cite{langson2004robust, goulart2006optimization} strategies. They induce different computational complexities and performances for the synthesized MPC controller, and a comprehensive comparison between these adaptive horizon strategies is beyond the scope of this paper. No matter which of the aforementioned adaptive horizon strategies is used, the proposed \augdist{} SLS MPC is guaranteed to be recursively feasible thanks to Theorem~\ref{thm:recursive_feasibility} and Lemma~\ref{lem:terminal_invariance}. 

In this work, we apply the adaptive horizon strategy described in~\eqref{eq:adaptive_horizon} to guarantee the ISS of the closed-loop system shown in the next section. In practice, we can keep running \augdist{} SLS MPC with a fixed horizon $T$ until infeasibility of problem~\eqref{eq:tightening} occurs when we decrease the horizon to $T-1$ and repeat the process. This will make sure that at each time step problem~\eqref{eq:tightening} looks ahead long enough for planning. In the next section, we prove the ISS of \augdist{} SLS MPC with horizons selected by the strategy~\eqref{eq:adaptive_horizon}.

\section{Input-to-state stability}
\label{sec:ISS}
In this section, we prove the input-to-state stability of the closed-loop system under the proposed MPC controller~\eqref{eq:mpc_policy} which adopts the adaptive horizon strategy~\eqref{eq:adaptive_horizon}. Recall that the closed-loop dynamics can be written as
\begin{equation} \label{eq:LTI_cl_dyn}
x(k+1) = \hat{A} x(k) + \hat{B} u_{MPC}^k(x(k)) + \eta(k) \text{ with } \eta(k) = \Delta_A x(k) + \Delta_B u_{MPC}^k(x(k)) + w(k).
\end{equation}
We first introduce the following definition of ISS.

\begin{definition}[Input-to-state stability, adapted from \cite{magni2006regional}] 
\label{def:ISS}	
Given a compact set $\mathcal{R} \subseteq \mathbb{R}^{n_x}$ including the origin in its interior, the system~\eqref{eq:LTI_cl_dyn} is said to be input-to-state stable in $\mathcal{R}$ if $\mathcal{R}$ is a robust forward invariant set for system~\eqref{eq:LTI_cl_dyn} and there exists a class-$\mathcal{KL}$ function $\beta$ and a class-$\mathcal{K}$ function $\gamma$ such that for any initial state $x(0) \in \mathcal{R}$ and bounded \augdists{} $\{ \eta(k) \}_{k=0}^\infty$ with $\lVert \eta(k) \rVert_\infty \leq C < \infty, k \geq 0$, the system state $x(k)$ exists for all $k \geq 0$ and satisfies 
\begin{equation}
\lVert x(k) \rVert_\infty \leq \beta(\lVert x(0) \rVert_\infty, k) + \gamma \big ( \sup_{0 \leq t \leq k} \lVert \eta(t) \rVert_\infty \big ).
\end{equation}
The set $\mathcal{R}$ is called the region of attraction (ROA) of system~\eqref{eq:LTI_cl_dyn}.
\end{definition}

Unlike the standard notion of ISS~\cite{sontag1989smooth, jiang2001input}, the ISS given in the above definition is with respect to the \augdists{} $\{\eta(k)\}_{k=0}^\infty$ where $\eta(k)$ includes both the exogenous disturbance input $w(k)$ and the uncertainty terms involving states $x(k)$. However, the implication of ISS in Definition~\ref{def:ISS} remains the same: (i) the origin is asymptotically stable for the closed-loop system when no uncertainty is present, i.e., when $\Delta_A = 0, \Delta_B = 0, \sigma_w = 0$; (ii) all the closed-loop trajectories are bounded when the \augdists{} are bounded; (iii) the closed-loop trajectory approaches the origin if $\eta(k) \rightarrow 0$ as $k \rightarrow \infty$. Next, we define the region of attraction (ROA) of the closed-loop system~\eqref{eq:LTI_cl_dyn} using the method proposed in~\cite{bujarbaruah2021simple}.

\begin{definition}[$N$-step robust controllable set]
For the MPC controller and the closed-loop dynamics~\eqref{eq:LTI_cl_dyn}, we define the $N$-step robust controllable set $\mathcal{C}_N$ to a given set $\mathcal{S}$ recursively as 
\begin{equation*}
	\mathcal{C}_0 = \mathcal{S}, \mathcal{C}_{k+1} = \text{Pre}(\mathcal{C}_k) \cap \mathcal{X}, \ k = 1, \cdots, N-1
\end{equation*}
where $\text{Pre}(\mathcal{S})$ is the preset of $\mathcal{S}$ defined as the set of states from which system~\eqref{eq:LTI_cl_dyn} evolve into $\mathcal{S}$ in one step for all possible realizations of model uncertainty and additive disturbances.
\end{definition}

\begin{definition}[ROA of the adaptive horizon robust MPC]
\label{def:ROA}
The region of attraction $\mathcal{R}$ for the proposed robust MPC method is defined as the union of the $N$-step robust controllable set to the terminal set $\mathcal{X}_T$ for $N \in \{0, \cdots, T_{max}\}$
\end{definition}

By the recursive feasibility of the proposed robust MPC method with the adaptive horizon strategy~\eqref{eq:adaptive_horizon}, we know that the ROA $\mathcal{R} \subseteq \mathcal{X}$ is robust forward invariant for the uncertain closed-loop dynamics~\eqref{eq:LTI_cl_dyn}. Therefore, the state of the closed-loop system~\eqref{eq:LTI_cl_dyn} is always bounded, and ISS additionally provides a qualitative description of the asymptotic behavior of the closed-loop system when the uncertainty parameters approach zero.

In standard MPC with a quadratic cost function and a fixed horizon, the value function of the finite time optimal control problem is continuous since it can be formulated as a parametric quadratic program~\cite{bemporad2002explicit}. Then, the continuity of the value function can be used to prove the ISS of the closed-loop system, as shown in~\cite{goulart2006optimization}. However, in our proposed robust MPC method, the value function, which we denote as $J^*_{MPC}(x(k))$, is not given by the solution to a parametric quadratic program; instead, we obtain it from
\begin{equation}\label{eq:value_function}
J^*_{MPC}(x(k)) = \min_{T \in \{1, \cdots, T_{max} \} } J^*_T(x(k)),
\end{equation}
and therefore cannot guarantee the continuity of $J^*_{MPC}(x(k))$. We can overcome this challenge using the results from~\cite{grune2014iss}, which states that the existence of a dissipative-form ISS-Lyapunov function verifies ISS of a discrete-time system under no regularity assumptions on the system dynamics or the ISS-Lyapunov function.  

\begin{definition}[ISS-Lyapunov function~\cite{grune2014iss}]
A function $V(x):\mathbb{R}^{n_x} \mapsto \mathbb{R}_\geq 0$ is called an ISS-Lyapunov function for system~\eqref{eq:LTI_cl_dyn} with a ROA $\mathcal{R}$ if there exist class-$\mathcal{K}_\infty$ functions $\alpha_1(\cdot), \alpha_2(\cdot), \alpha_3(\cdot)$, a class-$\mathcal{K}$ function $\sigma(\cdot)$ such that
\begin{subequations}
\begin{align}
&\alpha_1(\lVert x \rVert_\infty) \leq V(x) \leq \alpha_2(\lVert x \rVert_\infty), \quad \forall x \in \mathcal{R}  \label{eq:ISS_Lyap_cond_1}\\
&V(x(k+1)) - V(x(k)) \leq -\alpha_3(x(k)) + \sigma(\lVert \eta(k) \rVert_\infty), \quad \forall x(k) \in \mathcal{R}. \label{eq:ISS_Lyap_cond_2}
\end{align}
\end{subequations}
\end{definition}

As proved in~\cite[Theorem 2.3]{grune2014iss}, the closed-loop system~\eqref{eq:LTI_cl_dyn} is ISS if and only if there exists an ISS-Lyapunov function. We prove the ISS of system~\eqref{eq:LTI_cl_dyn} in Theorem~\ref{thm:ISS} by showing that the value function $J^*_{MPC}(x(k))$ is an ISS-Lyapunov function.

\begin{assumption} \label{assump:cost}
Let $\ell(x, u) = x^\top Q x + u^\top R u$ denote the stage cost in the nominal cost function~\eqref{eq:nominalcost}. The cost weights $Q$ and $R$ satisfy $Q \succ 0, R \succ 0$.
\end{assumption}

\begin{assumption} \label{assump:local_Lyapunov}
The terminal cost weight $Q_T$ in \eqref{eq:nominalcost} satisfies $Q_T \succ 0$ and 
\begin{equation}
x^\top (-Q_T + (Q + K^\top R K) + (\hat{A} + \hat{B}K)^\top Q_T (\hat{A} + \hat{B} K) )x \leq 0
\end{equation}
for all $x \in \mathcal{X}_T$ where $K$ is the local stabilizing controller defined in Assumption~\ref{assump:robust_invariant}.
\end{assumption}

With the given locally stabilizing controller $\kappa(x)=Kx$ satisfying Assumption~\ref{assump:robust_stable}, the terminal cost weight $Q_T$ satisfying Assumption~\ref{assump:local_Lyapunov} can be obtained by solving a discrete-time Lyapunov equation~\cite[Chapter 7]{borrelli2017predictive} for the linear system $x(k+1) = (\hat{A}+\hat{B}K)x(k)$. 

\begin{theorem} \label{thm:ISS}
Let Assumptions~\ref{assump:robust_stable}, \ref{assump:robust_invariant}, \ref{assump:cost}, \ref{assump:local_Lyapunov} hold and $x(0) \in \mathcal{R}$ where $\mathcal{R}$ denotes the ROA from Definition~\ref{def:ROA}. The value function $J^*_{MPC}(x(k))$ given in~\eqref{eq:value_function} is an ISS-Lyapunov function for the closed-loop system~\eqref{eq:LTI_cl_dyn} and therefore proves the ISS of system~\eqref{eq:LTI_cl_dyn}.
\end{theorem}

\begin{proof}
See Appendix~\ref{proof:ISS}.
\end{proof}

\section{Numerical examples}
\label{sec:simulation}
\subsection{Comparison to existing methods}
\label{sec:methods_comparison}
In this section, we compare \augdist{} SLS MPC with other robust MPC methods in the literature that can handle both model uncertainty and additive disturbances. Specifically, the following methods are considered for comparison:
\begin{enumerate}
	\item \emph{\augslsmpc{}}: Our proposed method which describes the effects of uncertainty through \augdists{} and synthesizes robust feedback controllers using SLS. 
	\item \emph{unif-df-MPC}: the method from~\cite{bujarbaruah2021simple} which uses a disturbance feedback approach based on a uniform over-approximation of the \augdist{}. 
	\item \emph{grid-SLS-MPC}: the SLS-based robust MPC method from~\cite{chen2020robust} which bounds the effects of uncertainty by grid-searching over a set of hyperparameters.
	\item \emph{tube-MPC}: the method proposed in~\cite{langson2004robust} which bounds the trajectories of system under uncertainty through a tube.
\end{enumerate}

\paragraph{\augslsmpc{} and unif-df-MPC}
Similar to \augslsmpc{}, unif-df-MPC also lumps model uncertainty and additive disturbances into a net-additive disturbance which we denote as the \augdist{} in this paper. However, unif-df-MPC ignores the dependence of the \augdists{} on the system states and control inputs; instead, it over-approximates all the \augdists{} by a single norm ball, i.e., by choosing a large-enough upper bound $\sigma$ such that $\lVert \eta_t \rVert_\infty \leq \sigma$ holds for all $x_t \in \mathcal{X}, u_t \in \mathcal{U}$. With this over-approximation, the disturbance feedback approach~\cite{goulart2006optimization} is applied in~\cite{bujarbaruah2021simple} to synthesize a robust controller. 

The method unif-df-MPC achieves good balance between simplicity and performance: indeed, its performance is comparable to tube-MPC in many examples shown in Section~\ref{sec:numerical_experiments}. The drawback is that the uniform norm upper bound $\sigma$ may be conservative. For example, when $x_0 = 0$, the actual \augdist{} $\eta_t$ are expected to be small since the system trajectory is near the origin. But with a large state constraint, e.g., $\mathcal{X} = \{x \vert \lVert x \rVert_\infty \leq 10\}$, the uniform norm upper bound $\sigma$ can be orders of magnitude larger than the true values of $\lVert\eta_t \rVert_\infty$ since all states in $\mathcal{X}$ are considered to derive the upper bound. In contrast, \augslsmpc{} does not suffer from this over-conservatism as it considers the dependence of $\eta_t$ on $(x_t, u_t, w_t)$ along the prediction horizon. Further, unif-df-MPC is a special case of \augslsmpc{} that sets $\sigma_t = \sigma, 0 \leq t \leq T-1$. Therefore, \augslsmpc{} is guaranteed to be less conservative than unif-df-MPC. 

\paragraph{\augslsmpc{} and grid-SLS-MPC}
Both methods use SLS to derive a convex inner approximation of the robust OCP but with different formulations. \augslsmpc{} relies on the dynamics of the \augdist{}~\eqref{eq:dyn_aug} to derive an inner approximation of the robust OCP~\eqref{eq:robustOCP} while grid-SLS-MPC directly exploits the mapping~\eqref{eq:response_true}. In grid-SLS-MPC, a tuple of three hyperparameters have to be grid-searched in order to make the robust OCP inner approximation convex. These hyperparameters can be interpreted as uniform upper bounds on the effects of the uncertainty along the prediction horizon. Although not directly comparable, \augslsmpc{} does not use uniform bounds for constraint tightening and the upper bounds $\sigma_t$ are treated as optimization variables instead of hyperparameters in the proposed inner approximation~\eqref{eq:tightening}. These features are desirable to derive more computationally efficient and less conservative robust optimal control formulations: this is confirmed by the numerical examples in Section~\ref{sec:numerical_experiments}. In~\cite{dean2019safely}, an SLS-based inner approximation similar to grid-SLS-MPC is proposed but is more conservative as shown in~\cite{chen2020robust}. Therefore, we do not compare our method with~\cite{dean2019safely}.

\paragraph{\augslsmpc{} and tube-MPC}
The tube-MPC method~\cite{langson2004robust} for handling both model uncertainty and additive disturbances requires first fixing the shape of a polytopic set, and then jointly synthesizing a tube and a feedback controller that constrains the system trajectories inside the tube by solving a convex program. The design of \augslsmpc{} can be interpreted in a similar fashion: we synthesize an LTV state feedback controller such that all trajectories are constrained in a tube characterized by a sequence of $\ell_\infty$ balls with varying sizes centered at the nominal predicted trajectory. However, the underlying feedback controllers are parameterized differently in tube-MPC and \augslsmpc{} which lead to different robust OCP formulations. We compare the conservatism and computational complexities of these two methods through numerical examples shown in the next subsection. Note that tube-MPC can handle polytopic model uncertainty which we leave for future research. 

\subsection{Experiments}
\label{sec:numerical_experiments}
We now compare the conservatism and computational complexity of the four robust MPC methods described above numerically. Our codes are publicly available at \url{https://github.com/ShaoruChen/Lumped-Uncertainty-SLS-MPC}.

\paragraph{Implementation details} For grid-SLS-MPC, we follow the implementation details described in~\cite{chen2020robust}. There are three hyperparameters in grid-SLS-MPC. In all the experiments, we use bisection~\cite[Algorithm 1]{chen2020robust} in the interval $[0.01 \ 100]$ to determine the lower and upper bounds on these hyperparameters and then apply a $5 \times 5 \times 5$ grid search between these bounds to look for a feasible tuple of hyperparameters. For tube-MPC, we choose the tube as the disturbance invariant set with an LQR controller. See~\cite{chen2020robust} for more details. All the experiments were implemented
in MATLAB R2019b with YALMIP~\cite{lofberg2004yalmip} and MOSEK~\cite{mosek} on an Intel i7-6700K CPU.

\subsubsection{Conservatism evaluation}
\label{sec:conservatism_evaluation}
\begin{figure}
	\centering
	\includegraphics[width = 0.6 \textwidth]{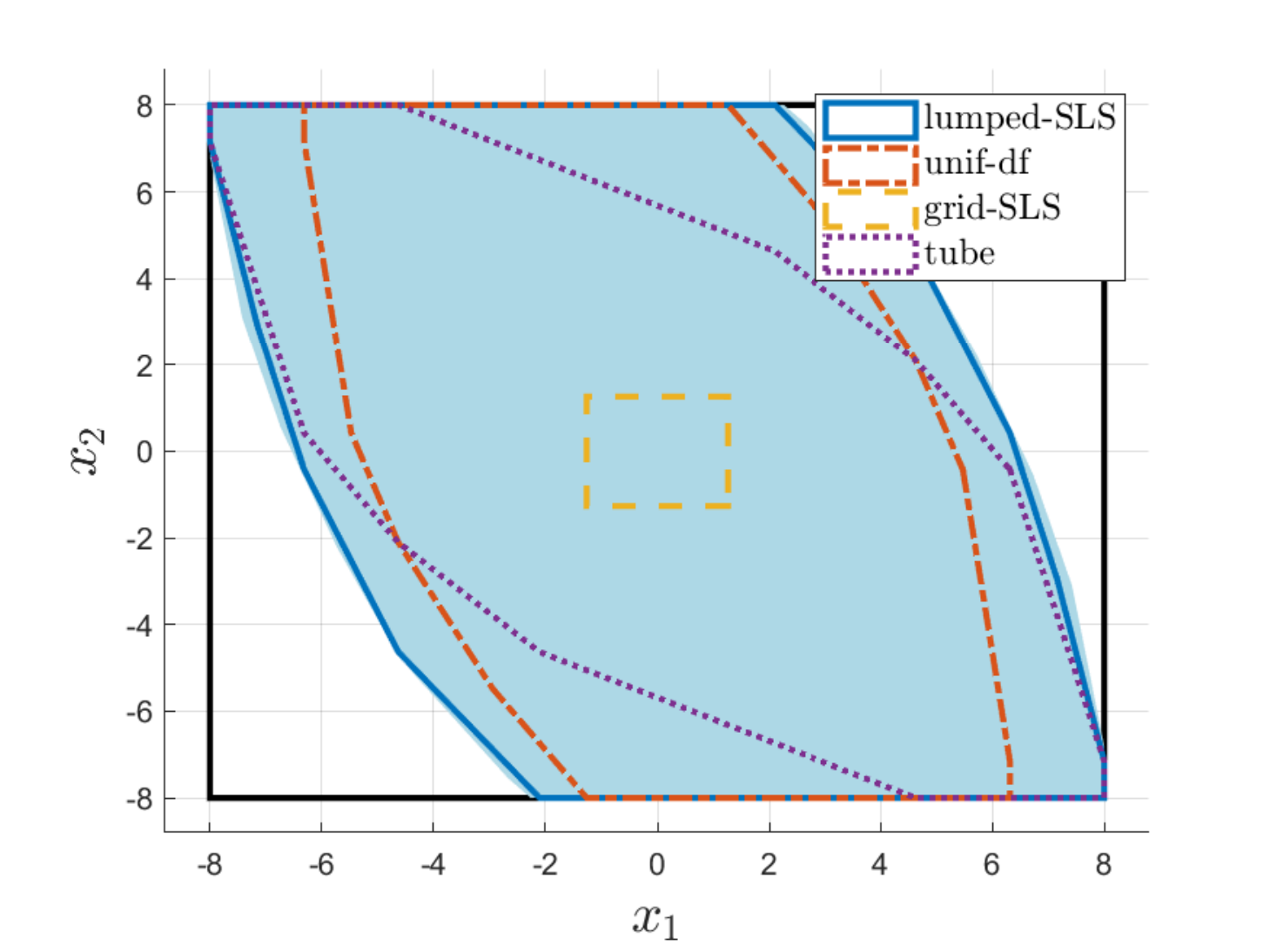}
	\caption{Conservatism evaluation of robust MPC methods. The shaded blue region denotes the maximum robust invariant set $\mathcal{X}_T$. The convex hull of the feasible initial conditions for each method is plotted.}
	\label{fig:conservatism}
\end{figure}

We use the two dimensional example from~\cite{bujarbaruah2020robust} to evaluate the performances of each robust MPC method. The nominal dynamics and the state and control input constraints are given as
\begin{equation} \label{eq:example}
\hat{A} = \begin{bmatrix}
1 & 0.15 \\ 0.1 & 1 
\end{bmatrix}, \quad \hat{B} = \begin{bmatrix}
0.1 \\ 1.1
\end{bmatrix}, \quad \mathcal{X} = \Big \{ x \in \mathbb{R}^2 \vert \begin{bmatrix}
-8 \\ -8
\end{bmatrix} \leq x \leq \begin{bmatrix}
8 \\8
\end{bmatrix}  \Big \}, \quad \mathcal{U} = \{ u \in \mathbb{R} \vert -4 \leq u \leq 4\}.
\end{equation}
We note that $\hat{A}$ is unstable. The model uncertainties $\{\Delta_A, \Delta_B\}$ are bounded by $\lVert \Delta_A\rVert_\infty \leq \epsilon_A= 0.1$, $\lVert \Delta_B \rVert_\infty \leq \epsilon_B = 0.1$. The additive disturbances satisfy $\lVert w(k) \rVert_\infty \leq 0.1$, i.e., $\sigma_w = 0.1$. We solve the robust OCP~\eqref{eq:robustOCP} with the terminal constraint $\mathcal{X}_T$ chosen as the maximal robust control invariant set (the shaded polytope in Figure~\ref{fig:conservatism}) found by the iterative algorithm~\cite[Algorithm 2]{grieder2003robust}. The MPC horizon is set to $T = 5$ and the cost weights are chosen as $Q = 10I, R = 1, Q_T = 10I$.

\paragraph{Comparison with $\mathcal{X}_T$}
Since the terminal set $\mathcal{X}_T$ is chosen as the maximal robust control invariant set, the robust OCP~\eqref{eq:robustOCP} is infeasible for any initial state $x_0 \notin \mathcal{X}_T$ and is feasible for all $x_0 \in \mathcal{X}_T$ with a robust piecewise affine controller as the solution. However, to find this robust piecewise affine controller we need to solve dynamic programming and multi-parametric quadratic programming problems with robustified constraints, which has an exponential complexity in the worst case~\cite{wang2009fast} and is considered intractable in practice. We can evaluate the conservatism of each robust MPC method by estimating their feasible regions (the set of $x_0$ for which the robust MPC method is feasible) and comparing them with the maximum robust control invariant set $\mathcal{X}_T$. 

To do so, we first sample initial conditions from a uniform grid of $288$ states within $\mathcal{X}_T$. Then we take the convex hull of all the feasible initial states of each method to estimate their feasible regions as shown in Figure~\ref{fig:conservatism}. We observe the following: (i) the feasible region of \augslsmpc{} almost matches the maximal robust control invariant set. In particular, \augslsmpc{} is feasible for all sampled initial states. This indicates that \augslsmpc{} is not conservative for the considered example, and furthermore only requires solving a computationally efficient convex quadratic program, in contrast to the dynamic programming approach; (ii) the feasible region of unif-df-MPC is a subset of that of \augslsmpc{}: as described in Section~\ref{sec:methods_comparison}, this is expected; (iii) grid-SLS-MPC gives the most conservative result in this example; (iv) the feasible regions of unif-df-MPC and tube-MPC are similar in area but cover different states. 

\paragraph{Comparison using varying uncertainty parameters}
We now stress test all four robust MPC methods for increasing uncertainty sizes. In the first test, we fix $\epsilon_B = 0.1, \sigma_w = 0.1$ and vary $\epsilon_A$ from $0.05$ to $0.25$ with step size $0.01$. In the second test, we fix $\epsilon_A = \epsilon_B = 0.1$ and vary $\sigma_w$ from $0.05$ to $0.8$ with step size $0.05$. For each of these tests, we sample $225$ uniformly-spaced states from the state constraint $\mathcal{X}$ and evaluate the feasibility of the robust MPC methods on these states. Since the maximal robust control invariant set $\mathcal{X}_T$ found in the previous subsection is no longer valid for the changed uncertainty parameters, we do not impose terminal constraints in these tests. 

\begin{figure}
\centering
\begin{subfigure}{0.49 \textwidth}
\includegraphics[width = \textwidth]{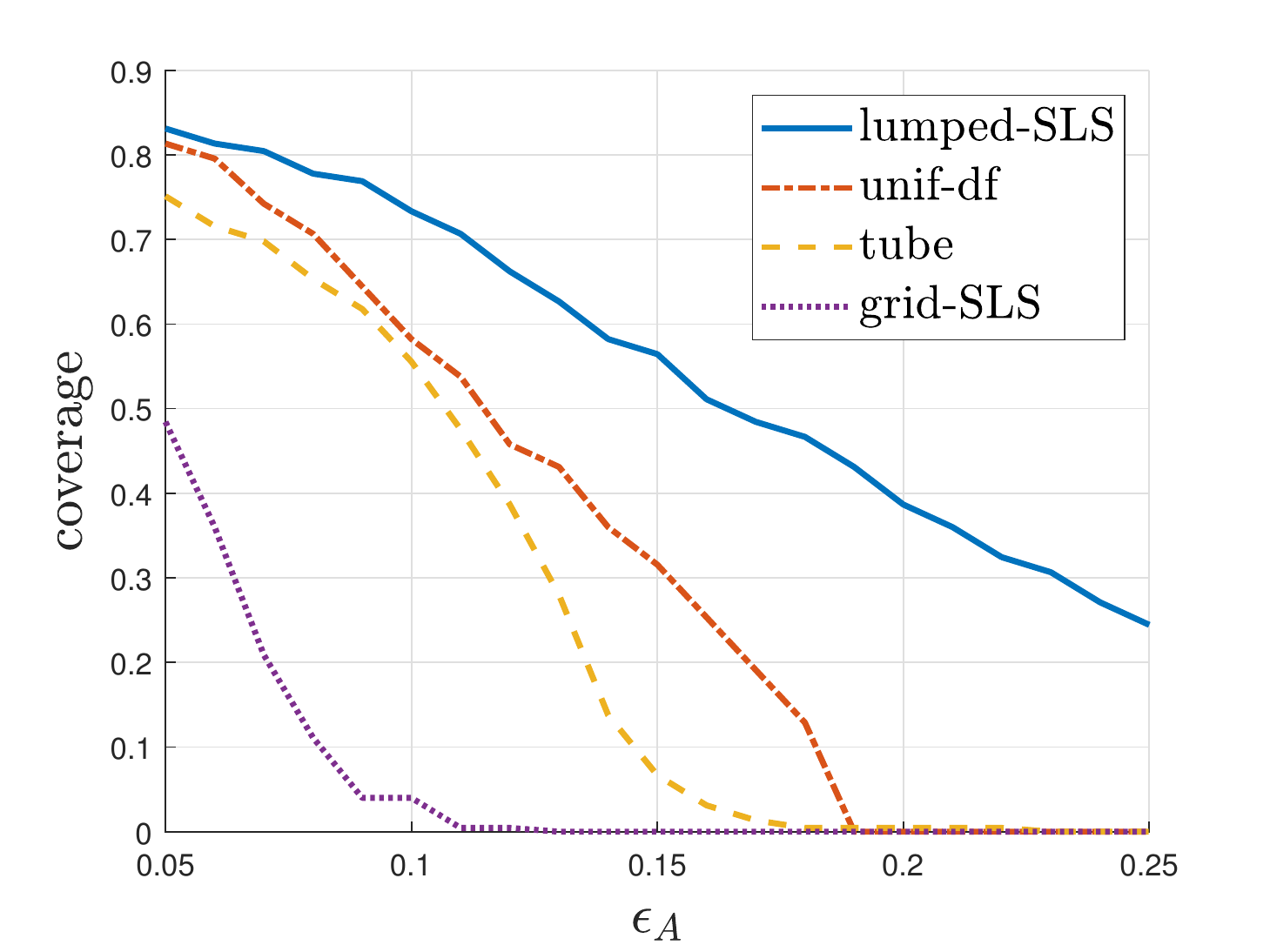}
\caption{Coverage of feasible regions for varying $\epsilon_A$ with $\epsilon_B = 0.1, \sigma_w = 0.1$}
\label{fig:coverage_eps_A}
\end{subfigure}
\hfill
\begin{subfigure}{0.49 \textwidth}
\includegraphics[width = \textwidth]{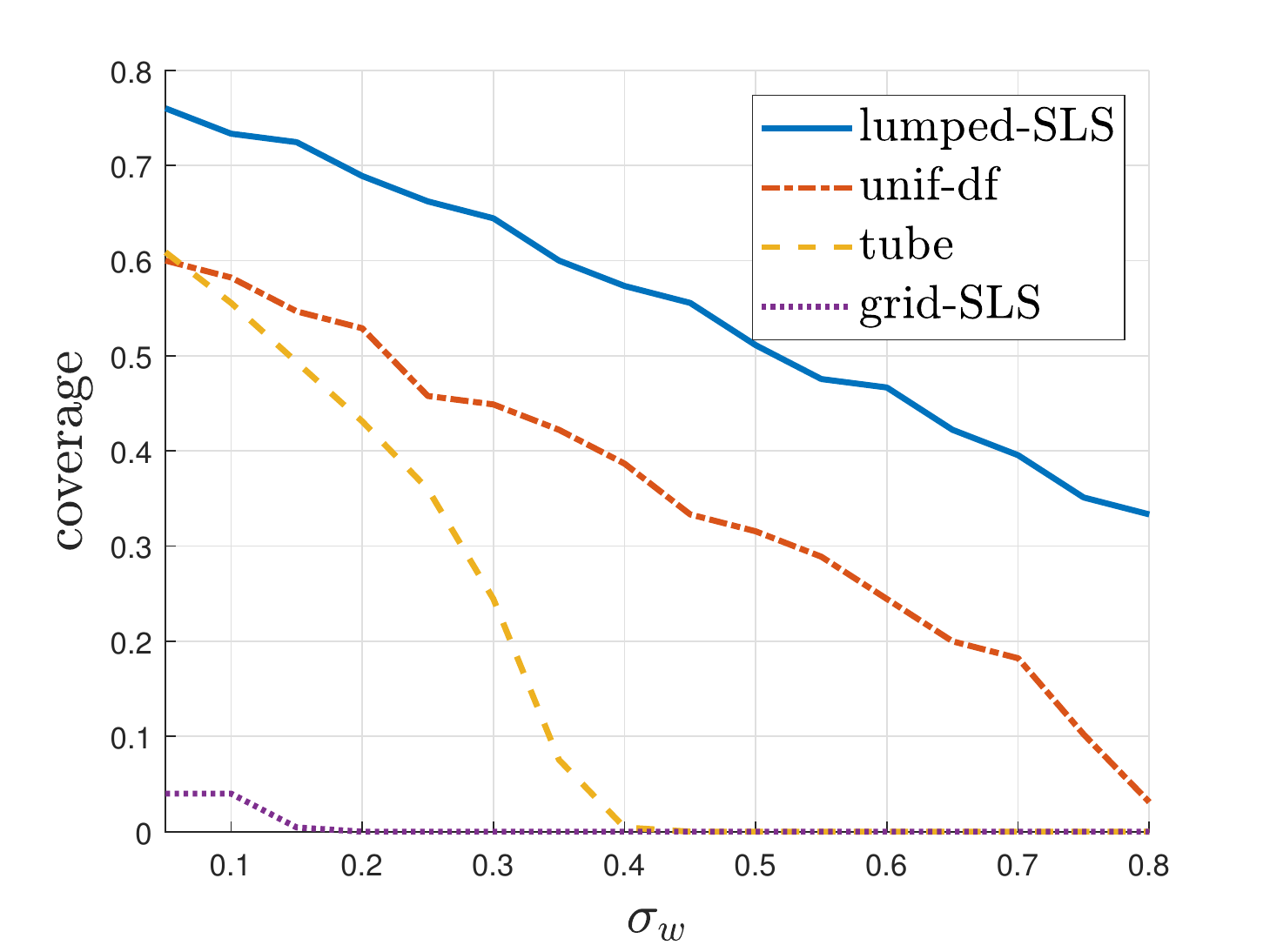}
\caption{Coverage of feasible regions for varying $\sigma_w$ with $\epsilon_A = 0.1, \epsilon_B = 0.1$.}
\label{fig:coverage_w}
\end{subfigure}
\caption{The coverage of the feasible regions (ratio of feasible sampled initial states) of the robust MPC methods with varying $\epsilon_A$ (Left) or $\sigma_w$ (Right). Our proposed method \augslsmpc{} consistently outperforms other candidate methods by a non-trivial margin.}
\label{fig:coverage}
\end{figure}

We evaluate the conservatism of each robust MPC method by its approximate feasible region coverage, which we define as the ratio of the feasible states to the sampled $225$ states. The results are shown in Fig.~\ref{fig:coverage}, and we observe that \augslsmpc{} outperforms all the other methods by a significant margin. Note that as the sampled states are from the state constraint $\mathcal{X}$ instead of the maximal robust invariant set for each tuple of uncertainty parameters $(\epsilon_A, \epsilon_B, \sigma_w)$, it is expected that the coverage of each robust MPC method decreases as the uncertainty parameter increases.

In Fig.~\ref{fig:coverage_eps_A}, \augslsmpc{} is shown to be the most resistant to the increasing level of model uncertainty since the slope of its coverage curve is the flattest. unif-df-MPC and tube-MPC achieves similar coverages with small $\epsilon_A$ but both of their coverages drop quickly as $\epsilon_A$ increases. grid-SLS-MPC is still the most conservative method for all cases. At $\epsilon_A = 0.1$, we recover the setup in the previous experiment, and confirm the observations (iii) and (iv) therein. In Fig.~\ref{fig:coverage_w}, \augslsmpc{} achieves the largest coverage for all values of $\sigma_w$ tested. Compared with Fig.~\ref{fig:coverage_eps_A}, the gap between unif-df-MPC and tube-MPC in coverage becomes even larger when we increase the norm of the additive disturbances. 


\begin{figure}
\centering 
\includegraphics[width = 0.5 \textwidth]{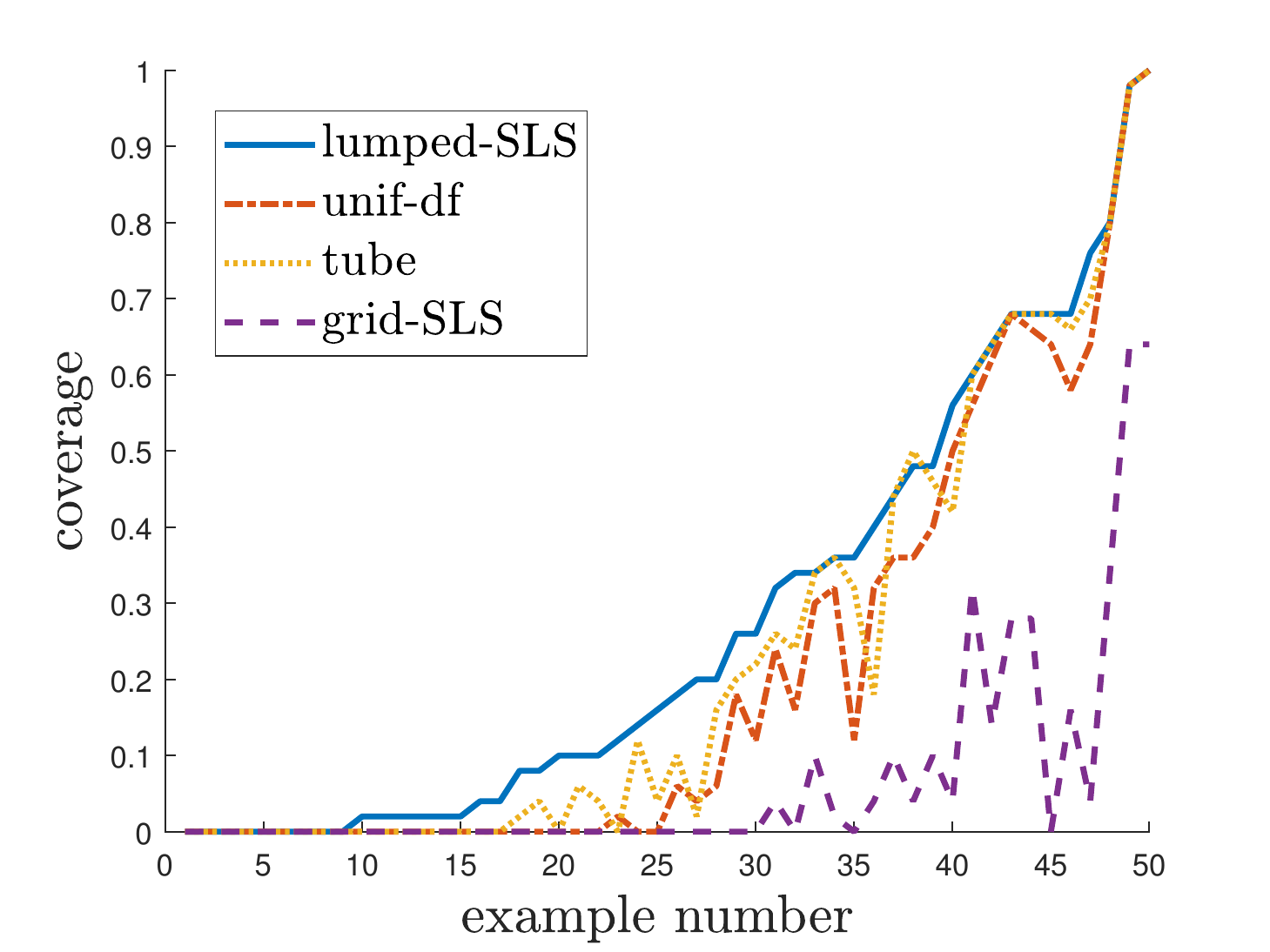} 
\caption{Coverage of each robust MPC method for $50$ randomly generated systems. Larger coverage indicates less conservatism. \augslsmpc{} outperforms all the other methods except at example No. $38$ where tube-MPC exceeds \augslsmpc{} by $2\%$ of coverage.}
\label{fig:random_coverage}
\end{figure}

\paragraph{Randomly generated systems}
We randomly generate $50$ two dimensional systems with nominal dynamics $\hat{A} \in \mathbb{R}^{2 \times 2}, \hat{B} \in \mathbb{R}^{2 \times 1}$. Each entry of $\hat{A}$ and $\hat{B}$ is uniformly sampled from the interval $[-2, 2]$ and $[-1, 1]$, respectively. The state and control input constraints are the same as in~\eqref{eq:example} with no terminal constraint imposed. The MPC horizon is set as $T = 5$ and the uncertainty parameters are chosen as $\epsilon_A = \epsilon_B = 0.1, \sigma_w = 0.1$. For each of the randomly generated system, we do a $10 \times 10$ uniform grid search over $\mathcal{X}$ and run the robust MPC methods on each sampled state. The coverage of each robust MPC method is computed as described above and is plotted in Fig.~\ref{fig:random_coverage} for each randomly generated system. We arrange the results in ascending order according to the coverage of \augslsmpc{}. Among the $50$ randomly generated systems, $25$ of them are open-loop unstable while the rest are open-loop stable. From Fig.~\ref{fig:random_coverage}, we observe that \augslsmpc{} has the largest coverage except at one example (No. $38$) where the coverage of tube-MPC exceeds that of \augslsmpc{} by $2\%$. 

\subsubsection{Computational complexity}
All four robust MPC methods solve a convex quadratic program at each MPC iteration, but they have different computational complexities. The number of variables in \augslsmpc{}, unif-df-MPC and grid-SLS-MPC are quadratic in the system dimension $n_x$ and the MPC horizon $T$ while being linear in $n_x$ and $T$ for tube-MPC. However, when considering $\ell_\infty \rightarrow \ell_\infty$ norm-bounded model uncertainty, the number of constraints in tube-MPC is linear in the horizon and exponential in the state and input dimensions since vertex enumeration is applied in tube-MPC to constrain the system trajectories inside the tubes. In contrast, the other three methods only tighten the constraints through norm-based inequalities: in these methods the number of constraints grows quadratically with the system dimension and the horizon. A detailed comparison is shown in~\cite[Table 1]{chen2020robust}. 

\begin{figure}
\centering
\begin{subfigure}[t]{0.49 \textwidth}
\includegraphics[width = \textwidth]{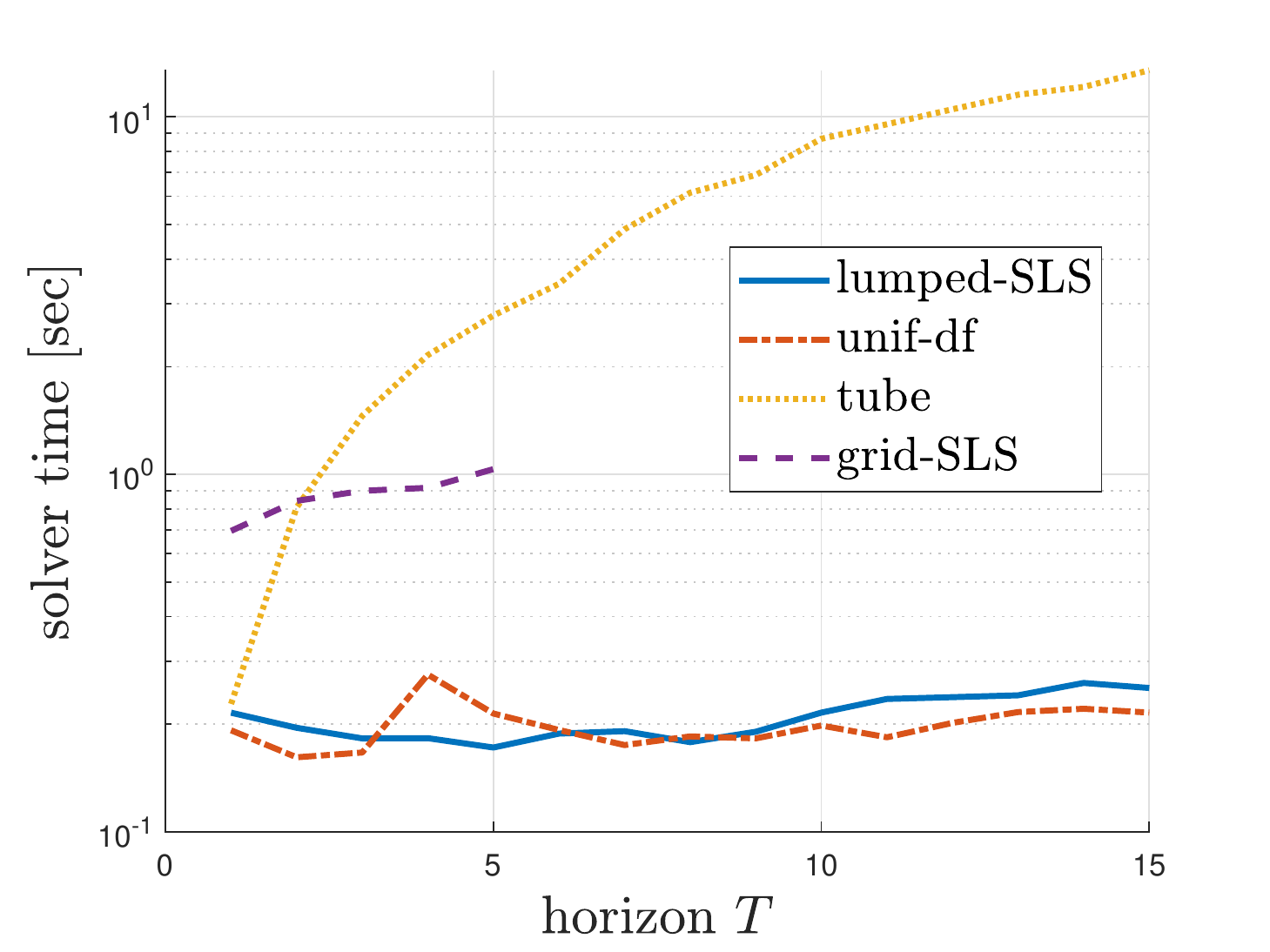}
\caption{Solver time of each robust MPC method for the numerical example considered in~\eqref{eq:example} with varying horizons.}
\label{fig:time_comparison}
\end{subfigure}
\hfill
\begin{subfigure}[t]{0.49 \textwidth}
\includegraphics[width = \textwidth]{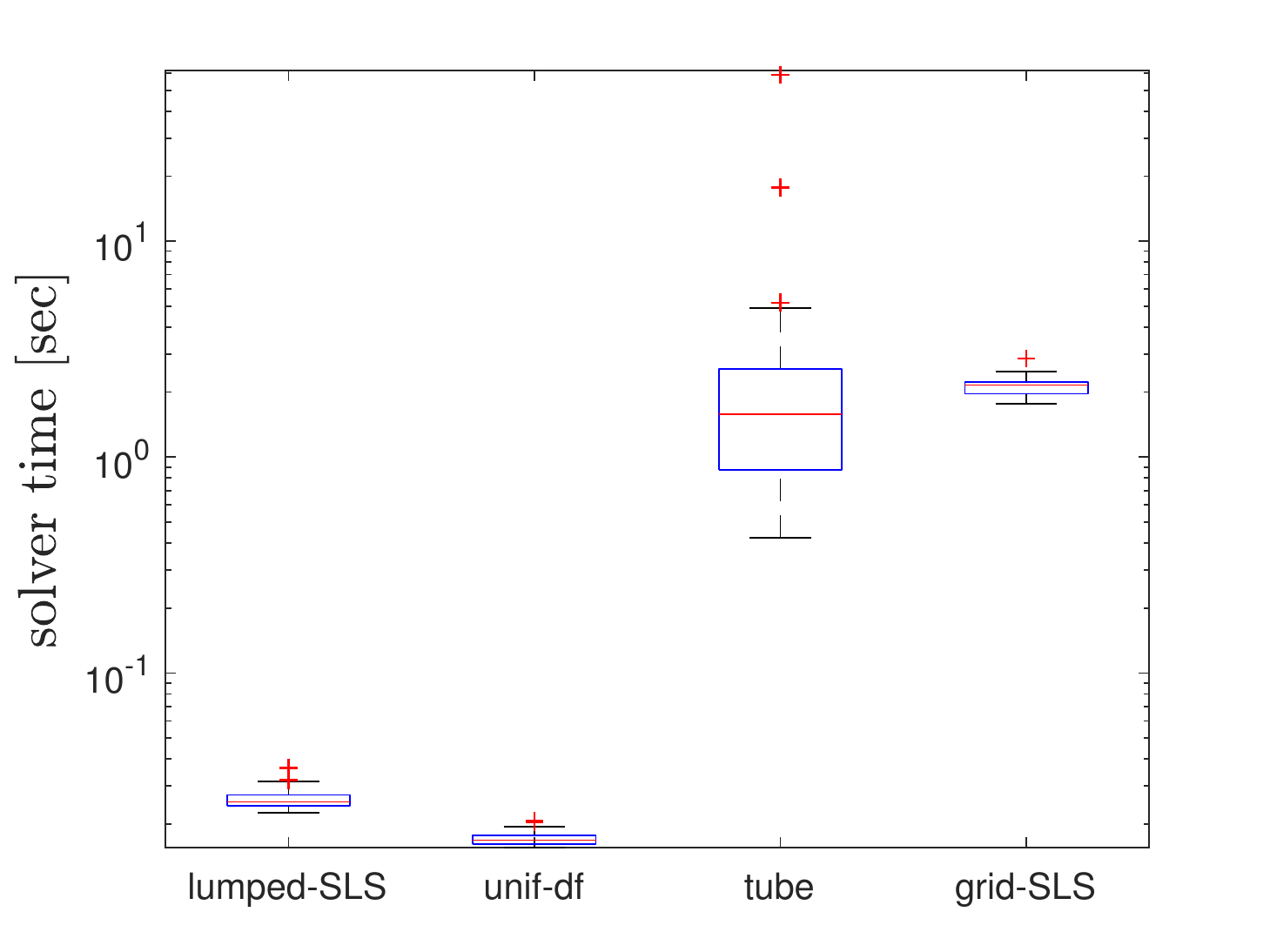}
\caption{Box plot of the solver time of each robust MPC method on a set of $100$ randomly generated systems. }
\label{fig:random_time_comparison}
\end{subfigure}
\caption{Left: With increasing horizion, grid-SLS-MPC becomes infeasible for horizon larger than $5$. The solver time of tube-MPC is significantly larger than that of \augslsmpc{} or unif-df-MPC, while the latter two methods achieve comparable computational time. Right: Statistics of the solver time of each robust MPC method on $100$ randomly generated systems with fixed horizon $T=10$. We observe that \augslsmpc{} and unif-df-MPC are much more computationally cheaper to run than the rest two methods.}
\end{figure}

In Fig.~\ref{fig:time_comparison}, we plot the solver time of each robust MPC method for the problem setup considered above with varying horizons. The initial condition is chosen as $x_0 = [1 \ 0]^\top$. The solver time of tube-MPC is significantly larger than that of \augslsmpc{} or unif-df-MPC mainly due to its large number of constraints in the quadratic programming formulation. grid-SLS-MPC becomes infeasible for horizon larger than $5$ and its solver time includes the bisection and grid search steps~\cite{chen2020robust}. Thus we see \augslsmpc{} and unif-df-MPC enjoy similar computational complexity.

Next, we randomly generate a hundred $2$-dimensional systems to compare the computational time. We generate $(\hat{A}, \hat{B})$ randomly as shown in Section~\ref{sec:conservatism_evaluation}. The robust OCP constraints are from~\eqref{eq:example} with no terminal constraints used, and the cost function is given by $Q = I, R = 1, Q_T = I$. For each sampled system $(\hat{A}, \hat{B})$, we fix the initial condition as $x_0 = [2 \ -1]^\top$ and the MPC horizon as $T = 10$. The uncertainty parameters are chosen as $\epsilon_A = \epsilon_B = \sigma_w = 0.05$. Then we apply all four robust MPC methods on the $100$ randomly generated system $(\hat{A}, \hat{B})$ ($82$ of them are open-loop unstable) and compare their solver time in Fig.~\ref{fig:random_time_comparison}. We mention that among the $100$ randomly generated systems, \augslsmpc{} is feasible for $55$ examples while unif-df-MPC tube-MPC, grid-SLS-MPC are feasible on $48$, $51$, $37$ examples, respectively. In Fig.~\ref{fig:random_time_comparison}, the solver time is reported only for feasible solutions of each robust MPC methods. We observe that \augslsmpc{} and unif-df-MPC are comparable, and both of them have much lower computational complexity than tube-MPC and grid-SLS-MPC. 

\section{Conclusion}
\label{sec:conclusion}
In this paper, we propose \augdist{} SLS MPC, a novel SLS-based robust MPC method for uncertain LTI systems subject to norm-bounded model uncertainty and additive disturbance. \augdist{} SLS MPC solves a convex inner approximation of the robust optimal control problem which guarantees the robust satisfaction of all state and control input constraints. It also enjoys recursive feasibility and input-to-state stability when combined with an adaptive horizon strategy. By jointly optimizing over a linear time-varying state feedback controller and the norm bounds of the state deviation in the prediction, our proposed method achieves significant improvement in conservatism compared with other baseline robust MPC methods, such as tube MPC, while being numerically efficient to solve. In future work, we will extend \augdist{} SLS MPC to accommodate other forms of model uncertainty and investigate output feedback MPC.

\bibliographystyle{ieeetr}
\bibliography{Refs}

\appendix

\section{Appendix}

\subsection{Proof of Lemma~\ref{lem:tightness}}
\label{proof:tightness}
With horizon $T = 1$, the state feedback controller is parameterized by $u_0 = K x_0$ with $K \in \mathbb{R}^{n_u \times n_x}$. The robust state constraint in the robust OCP~\eqref{eq:robustOCP} can be written as
\begin{equation} \label{eq:robust_state_constr}
f^\top (\hat{A} x_0 + \hat{B} K x_0 + \Delta_A x_0 + \Delta_B K x_0 + w_0) \leq b, \quad \forall \Delta_A \in \mathcal{P}_A, \Delta_B \in \mathcal{P}_B, w_0 \in \mathcal{W}
\end{equation}
for an arbitrary linear constraint $(f, b)\in \text{facet}(\mathcal{X})$. A tight upper bound on the left-hand side (LHS) of the above inequality is given by
\begin{equation} \label{eq:ub_tight}
\begin{aligned}
LHS & \leq f^\top (\hat{A}  + \hat{B} K)x_0 + \lVert f^\top \rVert_1 \lVert \Delta_A x_0 \rVert_\infty + \lVert f^\top \rVert_1 \lVert \Delta_B K x_0 \rVert_\infty + \lVert f^\top \rVert_1 \lVert w_0 \rVert_\infty  \\
& \leq f^\top (\hat{A}  + \hat{B} K)x_0 + \lVert f^\top \rVert_1 \lVert \Delta_A \rVert_\indinf \lVert x_0 \rVert_\infty + \lVert f^\top \rVert_1 \lVert \Delta_B \rVert_\indinf \rVert K x_0 \rVert_\infty + \lVert f^\top \rVert_1 \lVert w_0 \rVert_\infty  \\
& \leq f^\top (\hat{A}  + \hat{B} K)x_0 + \epsilon_A \lVert f^\top \rVert_1 \lVert x_0 \rVert_\infty + \epsilon_B \lVert f^\top \rVert_1 \rVert K x_0 \rVert_\infty + \lVert f^\top \rVert_1 \sigma_w
\end{aligned}
\end{equation}
by applying H\"{o}lder's inequality, the submultiplicativity of the $\ell_\infty$ norm, and the definition of the uncertainty set $\mathcal{P}_A, \mathcal{P}_B, \mathcal{W}$. It is easy to show that the upper bound given in~\eqref{eq:ub_tight} is achievable and thus tight. Therefore, the robust state constraint~\eqref{eq:robust_state_constr} is equivalent to 
\begin{equation} \label{eq:temp_1}
f^\top (\hat{A}  + \hat{B} K)x_0 +  \lVert f^\top \rVert_1 ( \epsilon_A\lVert x_0 \rVert_\infty + \epsilon_B \rVert K x_0 \rVert_\infty + \sigma_w) \leq b.
\end{equation}
We can similarly rewrite the robust control input and terminal constraints in this way. 

Now consider the robust OCP inner approximation~\eqref{eq:tightening}. For a given linear state constraint $(f, b) \in \text{facet}(\mathcal{X})$, the tightened constraint and the uncertainty norm upper bound constraint are given by
\begin{equation} \label{eq:temp_2}
\begin{aligned}
&f^\top (\hat{A}  + \hat{B} \widetilde{\Phi}_u^{0,0})x_0 + \lVert f^\top \widetilde{\Phi}_x^{0,0} \rVert_1 \sigma_0 \leq b, \\
&\epsilon_A \lVert \widetilde{\Phi}_x^{0,0} x_0 \rVert_\infty + \epsilon_B \rVert \widetilde{\Phi}_u^{0,0} x_0 \rVert_\infty + \sigma_w \leq \sigma_0.
\end{aligned}
\end{equation}
By the affine constraint~\eqref{eq:scaled_affine}, we have $\widetilde{\Phi}_x^{0,0} = I$ and $\widetilde{\Phi}_u^{0,0}$ is a free variable. As a result, the feasibility of \eqref{eq:temp_1} indicates feasibility of \eqref{eq:temp_2} by trivially matching $\widetilde{\Phi}_u^{0,0} = K, \sigma_0 = \epsilon_A \lVert \widetilde{\Phi}_x^{0,0} x_0 \rVert_\infty + \epsilon_B \rVert \widetilde{\Phi}_u^{0,0} x_0 \rVert_\infty + \sigma_w$. Since this relationship holds for all state, control input and terminal constraints, we have every feasible solution of the robust OCP~\eqref{eq:robustOCP} constructs a feasible one for~\eqref{eq:tightening}. Since~\eqref{eq:tightening} is an inner approximation of the robust OCP~\eqref{eq:robustOCP} by construction, we prove the equivalence between the convex inner approximation~\eqref{eq:tightening} and the robust OCP~\eqref{eq:robustOCP}.

\subsection{Proof of Theorem~\ref{thm:recursive_feasibility}}
\label{proof:feasibility}
We denote problem~\eqref{eq:tightening} at time $k$ as $\SLSOCP(k)$ with the initial state $x(k)$ and horizon $T_k  \in \{2, 3, \cdots, T_{max}\}$. Correspondingly, at time $k+1$, problem~\eqref{eq:tightening} is denoted as $\SLSOCP(k+1)$ with initial state $x(k+1)$ and horizon $T_{k+1}$. We want to show that when $\SLSOCP(k)$ is feasible with horizon $T_k$, then $\SLSOCP(k+1)$ is also feasible with horizon $T_{k+1} = T_k -1$. In this proof, notations with subscripts $k$ and $k+1$, such as $\tildePhixk$ and $\tildePhixkp$, refer to those used in problems $\SLSOCP(k)$ and $\SLSOCP(k+1)$, respectively. 

\paragraph{Sketch of the proof} Assume $\SLSOCP(k)$ is feasible with a feasible solution $(\tildePhixk, \tildePhiuk, \{\sigma_{i,k} \}_{i=0}^{T_k})$. Then, the synthesized linear state feedback controller $\KK_k = \tildePhiuk \tildePhixk^{-1}$ is a robust one for the robust OCP~\eqref{eq:robustOCP} at time $k$ and we denote this control policy by
\begin{equation}\label{eq:robust_policy_k}
U_k = \{ u_{k\vert k}^*, u_{k+1\vert k}^*(\cdot), \cdots, u_{k+T_k-1 \vert k}^*(\cdot)  \}
\end{equation}
We use $x_{t\vert k}$ and $u_{t \vert k}(\cdot)$ to denote the predicted state and control policy at time $t$ in $\SLSOCP(k)$. Note that $U_k$ is just a different representation of the controller $\KK_k$ with the correspondence $u_{t\vert k}^*(x_{k:t \vert k}) = \sum_{i=0}^{t-k} K_k^{t-k, t-k-i}x_{k+i \vert k}$. Then, at time $k+1$, a candidate robust feedback policy for the robust OCP~\eqref{eq:robustOCP} is given by
\begin{equation}\label{eq:robust_policy_kp}
U_{k+1} = \{ u_{k+1\vert k}^*(\cdot), \cdots, u_{k+T_{k+1} \vert k}^*(\cdot)  \}
\end{equation}
with horizon $T_{k+1}= T_k-1$ which truncates~\eqref{eq:robust_policy_k}. However, since problem~\eqref{eq:tightening} is only an inner approximation of the robust OCP~\eqref{eq:robustOCP}, whether the truncated policy $U_{k+1}$ will generate a feasible solution to the inner approximation~\eqref{eq:tightening} in the space of system responses remains to be shown.

In this proof, we resolve this issue in three steps. The first step is to recover the robust linear state feedback controller $\KK_{k+1}$ from the truncated policy sequence $U_{k+1}$ and derive the corresponding system responses $\{\tildePhixkp, \tildePhiukp \}$. In the second step, we illustrate the relationship between the constructed $\{\tildePhixkp, \tildePhiukp \}$ and the existing feasible solution $\{\tildePhixk, \tildePhiuk \}$ to $\SLSOCP(k)$. By exploiting the connection between these two sets of system responses, in the third step, we verify that $\{\tildePhixkp, \tildePhiukp \}$ together with the generated norm upper bound sequence $\{\sigma_{i,k+1} \}_{i=0}^{T_{k+1}}$ is indeed a feasible solution to $\SLSOCP(k+1)$, hence proving the feasibility of $\SLSOCP(k+1)$.

\paragraph{Step 1: candidate solution construction}
Recall that we use $x_{t\vert k}, u_{t \vert k}(\cdot)$ to denote the predicted state and control policy at time $t$ with $k \leq t \leq k + T_k$ in $\SLSOCP(k)$ where the current state is $x_{k\vert k} = x(k)$. The terms $x_{t \vert k+1}, u_{t \vert k+1}(\cdot)$ for $k+1 \leq t \leq k+1+T_{k+1}$ are defined similarly. When $\SLSOCP(k)$ is feasible, its solution $\{\tildePhixk, \tildePhiuk, \{ \sigma_{i,k}\}_{i=0}^{T_k} \}$ generates a robust state feedback controller $\KK_k = \tildePhiuk \tildePhixk^{-1} \in \mathcal{L}_{TV}^{T_k}$ which gives $u_{t \vert k}(x_{k:t \vert k}) = \sum_{i=0}^{t-k} K_{k}^{t-k,t-k-i} x_{k+i \vert k}$ for $k \leq t \leq k+T_k$. We let $\Sigma_k$ be the block diagonal matrix consisting of the norm upper bounds $\{ \sigma_{i,k}\}_{i=0}^{T_k}$ as shown in~\eqref{eq:tightening}, and $\{ \Phixk, \Phiuk \}$ be the corresponding system responses directly acting on the \augdists{} (see~\eqref{eq:system_response_Phi}). As shown in Remark~\ref{remark:affine_constr}, we know that $\Phixk = \tildePhixk \Sigma_k^{-1}$, $\Phiuk = \tildePhiuk \Sigma_k^{-1}$. We choose $T_{k+1} = T_k -1$ as the horizon of $\SLSOCP(k+1)$.

At time $k+1$, both $x(k)$ and $x(k+1)$ are known. The truncated policy~\eqref{eq:robust_policy_kp} generates a state feedback controller $\KK_{k+1} \in \mathcal{L}_{TV}^{T_{k+1}}$ which satisfies
\begin{equation} \label{eq:shift_controller}
u_{t \vert k+1}(x_{k+1:t \vert k+1}) = u_{t \vert k}(x_{k:t \vert k}), \quad k+1 \leq t \leq k+T_k-1
\end{equation}
with $x_{k \vert k} = x(k), x_{k+1 \vert k} = x_{k+1 \vert k+1} = x(k+1)$. Eqn.~\eqref{eq:shift_controller} says that $u_{t \vert k+1}(\cdot)$ returns the same control input as $u_{t \vert k}(\cdot)$ given the same predicted trajectory. With the parameterization of $\KK_k$ and $\KK_{k+1}$, Eqn.~\eqref{eq:shift_controller} is equivalent to
\begin{equation} 
\begin{aligned}
& K_{k+1}^{0,0} x_{k+1 \vert k+1} = K_k^{1,1} x_{k \vert k} + K_k^{1, 0} x_{k+1 \vert k+1} \\
& K_{k+1}^{1,1} x_{k +1 \vert k+1} + K_{k+1}^{1,0} x_{k +2 \vert k+1} = K_k^{2,2} x_{k \vert k} + K_k^{2,1} x_{k +1 \vert k+1} + K_k^{2,0} x_{k +2 \vert k+1} \\
& \cdots \\
& K_{k+1}^{T_{k+1}-1, T_{k+1}-1} x_{k +1 \vert k+1} + \cdots + K_{k+1}^{T_{k+1}-1,0} x_{k +T_{k+1} \vert k+1} = \\
& K_k^{T_k-1, T_k-1} x_{k\vert k} + K_k^{T_k-1, T_k-2} x_{k+1 \vert k+1} + K_k^{T_k-1, T_k-3} x_{k+2 \vert k+1} + \cdots + K_k^{T_k-1, 0} x_{k+T_k -1\vert k+1} \\
\end{aligned}
\end{equation}
for all possible $x_{t \vert k+1}, k+2 \leq t \leq k + T_k -1$. This immediately leads to 
\begin{subequations} \label{eq:K_eqs}
	\begin{align}
	\begin{split} \label{eq:replicate_1}
	K_{k+1}^{i,j} & = K_{k}^{i+1,j},\ \text{for all } i = 0, 1, \cdots, T_{k+1}, j = 0, 1, \cdots, i-1 
	\end{split} \\
	\begin{split}\label{eq:replicate_2}
	K_{k+1}^{i, i} x_{k+1 \vert k+1} & = K_{k}^{i+1, i+1} x_{k\vert k} + K_{k}^{i+1, i} x_{k+1\vert k+1}, \ \text{for all } i = 0, 1, \cdots, T_{k+1}
	\end{split}
	\end{align}
\end{subequations}
Under the condition $x_{k+1 \vert k+1} = x(k+1) \neq 0$\footnotemark, the system of linear equations~\eqref{eq:K_eqs} always has a feasible solution $\KK_{k+1}$ which is our candidate robust linear state feedback controller realizing the truncated policy~\eqref{eq:robust_policy_kp}. By~\eqref{eq:system_response_K}, a candidate system response solution $\{\Phixkp, \Phiukp\}$ acting on the \augdists{} can be explicitly constructed from $\KK_{k+1}$ by
\footnotetext{We can provide non-zero nominal control inputs $\KK_{k+1}(:, 0)x(k+1)$ on the left-hand side of~\eqref{eq:replicate_2} only when $x(k+1)$ is non-zero. In this case the state feedback controller $\uu = \KK\xx$ is equivalent to an affine feedback controller used in~\cite{goulart2006optimization}. We can remove the non-zero state assumption by augmenting system~\eqref{eq:uncertain_system} with state $\bar{x} = [x; 1]$.}
\begin{equation} \label{eq:Phixu_construction}
\Phixkp = (I - Z(\hat{\sA} + \hat{\sB} \KK_{k+1}))^{-1}, \quad \Phiukp = \KK_{k+1} (I - Z( \hat{\sA} + \hat{\sB} \KK_{k+1}))^{-1}
\end{equation}
where the size of block diagonal matrices $\hat{\sA}, \hat{\sB}$ are adapted according to the horizon $T_{k+1}$. We choose the candidate solutions of the uncertainty norm upper bounds $\{\sigma_{i,k+1}\}_{i=0}^{T_{k+1}}$ for $\SLSOCP(k+1)$ as $\sigma_{i,k+1} = \sigma_{i+1, k}$ for $0 \leq i \leq T_{k+1}$, and $\Sigma_{k+1}$ be the corresponding block diagonal matrix. Then, we can construct a candidate solution $\{ \tildePhixkp, \tildePhiukp \}$ by letting $\tildePhixkp = \Phixkp \Sigma_{k+1}$, $\tildePhiukp = \Phiukp \Sigma_{k+1}$ for $\SLSOCP(k+1)$. Before we show the feasibility of the constructed candidate solution $\{ \tildePhixkp, \tildePhiukp, \{\sigma_{i,k+1} \}_{i=0}^{T_{k+1}} \}$, the relationship between $\{\Phixkp, \Phiukp\}$ and $\{\Phixk, \Phiuk\}$ has to be established.

\paragraph{Step 2: properties of the candidate solution}
At time $k+1$, we denote the \augdist{} at time $k$ by $\eta_{k \vert k} = x(k+1) - x(k)$. Since we use the truncated control policy at time $k+1$ as shown in~\eqref{eq:shift_controller}, we have that
\begin{equation}\label{eq:matching}
x_{t \vert k+1} = x_{t \vert k}, \quad u_{t \vert k+1} = u_{t \vert k}, \quad t = k+1, \cdots, k + T_{k+1}
\end{equation}
under the controller $\KK_{k+1}$ in the case that the same predicted uncertainty parameters are given from time $k$ to time $k+T_{k+1}$, i.e., we consider the case when $\Delta_{A \vert k+1} = \Delta_{A \vert k}, \Delta_{B \vert k+1} = \Delta_{B \vert k }, w_{t \vert k+1} = w_{t \vert k}$ for $k+1 \leq t \leq k+T_{k+1}$, where $\Delta_{A \vert k}, \Delta_{B \vert k }$ denote the predicted model uncertainty parameters and $w_{t \vert k}$ denote the predicted disturbance at time $t$ in the $k$-step of MPC. Now we consider different realizations of the uncertainty parameters to explore the connection between $\{\Phixkp, \Phiukp\}$ and $\{\Phixk, \Phiuk\}$. 

First, we let all predicted model uncertainty parameters $\Delta_{A \vert k+1}, \Delta_{B \vert k+1}$ and $w_{t \vert k+1}, k+1 \leq t \leq k + T_{k+1}$ be zero and so are the \augdists{}. By the mapping of system responses~\eqref{eq:system_response_Phi} and the matching constraint~\eqref{eq:matching}, we have 
\begin{equation}\label{eq:new_sol_map_1}
\begin{aligned}
\PPhi_{x,k+1}^{i,i} x_{k+1 \vert k+1} = \PPhi_{x,k}^{i+1,i+1} x_{k \vert k} + \PPhi_{x,k}^{i+1,i} \bar{w}_{k \vert k} \\
\PPhi_{u,k+1}^{i,i} x_{k+1 \vert k+1} = \PPhi_{u,k}^{i+1,i+1} x_{k \vert k} + \PPhi_{u,k}^{i+1,i} \bar{w}_{k \vert k}
\end{aligned}
\end{equation}
for $i = 0, 1, \cdots, T_{k+1}$. Next we let $\Delta_{A \vert k+1} = 0, \Delta_{B \vert k+1} = 0$, $w_{t \vert k+1} = 0$ for $t \neq k+1$, but allow $w_{t \vert k+1}$ to be non-zero at $t = k+1$. Then the \augdists{} are all zero except $\bar{w}_{k+1 \vert k+1}$, and $\bar{w}_{k+1 \vert k+1} = w_{k+1 \vert k+1}$. Note that we apply the same predicted uncertainty parameters in $\SLSOCP(k)$ such that $\bar{w}_{t \vert k} = \bar{w}_{t \vert k+1}$ for $k+1 \leq t \leq k+T_{k+1}$. By the matching constraint~\eqref{eq:matching} and the system responses map~\eqref{eq:system_response_Phi}, we have
\begin{equation}\label{eq:sol_k_1}
\begin{aligned}
\PPhi_{x,k+1}^{i,i} x_{k+1 \vert k+1} + \PPhi_{x,k+1}^{i, i-1} \eta_{k+1 \vert k+1} = \PPhi_{x,k}^{i+1,i+1} x_{k \vert k} + \PPhi_{x,k}^{i+1,i} \eta_{k \vert k} + \PPhi_{x,k}^{i+1, i-1} \eta_{k+1 \vert k}\\
\PPhi_{u,k+1}^{i,i} x_{k+1 \vert k+1} + \PPhi_{u,k+1}^{i,i-1} \eta_{k+1 \vert k+1} = \PPhi_{u,k}^{i+1,i+1} x_{k \vert k} + \PPhi_{u,k}^{i+1,i} \eta_{k \vert k} + \PPhi_{u,k}^{i+1, i-1} \eta_{k+1 \vert k}
\end{aligned}
\end{equation}
for $i = 1, \cdots, T_{k+1}$. Subtracting Eqn.~\eqref{eq:new_sol_map_1} from both sides of~\eqref{eq:sol_k_1}, we conclude that $\PPhi_{x,k+1}^{i, i-1} = \PPhi_{x, k}^{i+1, i-1}, \PPhi_{u,k+1}^{i, i-1} = \PPhi_{u, k}^{i+1, i-1}$ for $i = 1, \cdots, T_{k+1}$ since the predicted \augdist{} $\eta_{k+1 \vert k+1} = \eta_{k+1 \vert k}$ can be arbitrary. Similarly, by letting $w_{t \vert k+1}$ be non-zero at time $t = k+2$ and zero otherwise, we can prove $\PPhi_{x,k+1}^{i, i-2} = \PPhi_{x, k}^{i+1, i-2}, \PPhi_{u,k+1}^{i, i-2} = \PPhi_{u, k}^{i+1, i-2}$ for $i = 2, \cdots, T_{k+1}$. Repeat this process and we have
\begin{equation} \label{eq:new_sol_map_2}
\PPhi_{x,k+1}^{i, j} = \PPhi_{x, k}^{i+1, j}, 
\quad \PPhi_{u,k+1}^{i, j} = \PPhi_{u, k}^{i+1, j}, \quad \text{ for } i = 1, 2, \cdots, T_{k+1}, j = 0, \cdots, i-1.
\end{equation}
Eqn.~\eqref{eq:new_sol_map_1} and \eqref{eq:new_sol_map_2} describe the relationship between $\{\Phixkp, \Phiukp\}$ and $\{\Phixk, \Phiuk\}$. Next, we will show that our proposed candidate solution $\{ \tildePhixkp, \tildePhiukp, \{\sigma_{i,k+1} \}_{i=0}^{T_{k+1}} \}$ is feasible for $\SLSOCP(k+1)$.

\paragraph{Step 3: feasibility of the candidate solution} Recall that $\tildePhixkp = \Phixkp \Sigma_{k+1}$, $\tildePhiukp = \Phiukp \Sigma_{k+1}$. Since $\{\Phixkp, \Phiukp\}$ in~\eqref{eq:Phixu_construction} satisfies the affine constraint~\eqref{eq:nominal_affine} by construction, we have $\{ \tildePhixkp, \tildePhiukp, \{\sigma_{i,k+1} \}_{i=0}^{T_{k+1}} \}$ satisfy the affine constraint in~\eqref{eq:tightening}. 

For the tightened constraints~\eqref{eq:state_tightening}, \eqref{eq:terminal_tightening}, \eqref{eq:input_tightening}, we take the state constraint tightening~\eqref{eq:state_tightening} as an example for analysis. For an arbitrary linear constraint $(f, b) \in \text{facet}(\mathcal{X})$, we have
\begin{align*}
&f^\top \widetilde{\Phi}_{x,k+1}^{t,t} x_{k+1 \vert k+1} + \sum_{i=1}^t \lVert f^\top \widetilde{\Phi}_{x,k+1}^{t,t-i} \rVert_1 \\
= &f^\top \Phi_{x,k+1}^{t,t} x_{k+1 \vert k+1} + \sum_{i=1}^t \lVert f^\top \Phi_{x,k+1}^{t,t-i} \rVert_1 \sigma_{i-1,k+1} \quad (\text{by } \tildePhixkp = \Phixkp \Sigma_{k+1})\\
=&  f^\top (\Phi_{x,k}^{t+1,t+1} x_{k \vert k} + \Phi_{x,k}^{t+1, t} \eta_{k\vert k}) + \sum_{i=1}^t  \lVert f^\top \Phi_{x,k}^{t+1,t-i} \rVert_1 \sigma_{i,k} \quad (\text{by Eqn.\eqref{eq:new_sol_map_1} and \eqref{eq:new_sol_map_2}}) \\
\leq &  f^\top \Phi_{x,k}^{t+1,t+1} x_{k \vert k} + \lVert f^\top \Phi_{x,k}^{t+1, t} \rVert_1 \lVert \eta_{k\vert k}  \rVert_\infty + \sum_{i=1}^t  \lVert f^\top \Phi_{x,k}^{t+1,t-i} \rVert_1 \sigma_{i,k} \quad  (\text{by H\"older's inequality})\\
\leq &  f^\top \Phi_{x,k}^{t+1,t+1} x_{k \vert k} + \sum_{i=1}^{t+1} \lVert f^\top \Phi_{x,k}^{t+1,t+1-i} \rVert_1 \sigma_{i-1,k} \quad (\text{by the upper bound } \lVert \bar{w}_{k\vert k} \rVert_\infty \leq \sigma_{0,k})\\
= & f^\top \widetilde{\Phi}_{x,k}^{t+1,t+1} x_{k \vert k} + \sum_{i=1}^{t+1} \lVert f^\top \widetilde{\Phi}_{x,k}^{t+1,t+1-i} \rVert_1  \quad (\text{by } \tildePhixk = \Phixk \Sigma_k ) \\
\leq & b \quad (\text{by the feasibility of } \SLSOCP(k))
\end{align*}
for $t =0, 1, \cdots, T_{k+1}$. Therefore, our constructed candidate solution satisfy all the tightened state constraints in~\eqref{eq:tightening} and we can prove the feasibility of the constructed solution for the tightened terminal state~\eqref{eq:terminal_tightening} and control input constraint~\eqref{eq:input_tightening} similarly. It is also straight forward to apply the above techniques to prove the the feasibility of $\{\tildePhixkp, \tildePhiukp, \{\sigma_{i, k+1}\}_{i=0}^{T_{k+1}} \}$ for the upper bound constraint~\eqref{eq:suff_cond_ub}. Hence, we prove that our constructed solution $\{\tildePhixkp, \tildePhiukp, \{\sigma_{i, k+1}\}_{i=0}^{T_{k+1}} \}$ is feasible for $\SLSOCP(k+1)$.

\subsection{Proof of Theorem~\ref{thm:ISS}}
\label{proof:ISS}
By Assumption~\ref{assump:cost}, the stage cost $\ell(x, u) = x^\top Q x + u^\top R u$ is positive definite and there exists a class-$\mathcal{K}_\infty$ function $\alpha_1(\cdot)$ such that $\alpha_1(\lVert x \rVert_\infty) \leq \ell(x, 0) \leq J^*_{MPC}(x)$ for all $x \in \mathcal{R}$ where $\mathcal{R}$ is the region of attraction given in Definition~\ref{def:ROA}. Using an argument similar to~\cite[Proposition 17]{goulart2006optimization}, we can show that each $J^*_N(x)$ for $N \in \{1, \cdots, T_{max}\}$ is a strictly convex, piecewise quadratic function in $x$. Since the value function $J^*_{MPC}(x)$ is the point-wise minimum of $J^*_N(x)$ for $N = 1, \cdots, T_{max}$, there exists a class-$\mathcal{K}_\infty$ function $\alpha_2(\cdot)$ such that $J^*_{MPC}(x) \leq \alpha_2(\lVert x \rVert_\infty)$ for all $x \in \mathcal{R}$. Recall that $T_k$ is the horizon of the MPC at time $k$ which achieves the minimum nominal cost according to the adaptive horizon strategy~\eqref{eq:adaptive_horizon}. By the recursive feasibility of our proposed MPC method, the ROA $\mathcal{R}$ is a robust forward invariant set for the closed-loop system. To show $J^*_{MPC}(x(k))$ satisfies condition~\eqref{eq:ISS_Lyap_cond_2}, we divide our discussion into two cases:

\noindent \textbf{Case 1 $T_k =1$}: When $T_k = 1$, by Assumption~\ref{assump:local_Lyapunov} we have
\begin{equation*}
\begin{aligned}
J^*_{MPC}(x(k)) &= \ell(\bar{x}^*_{k \vert k}, \bar{u}^*_{k \vert k}) + \bar{x}^{*,\top}_{k+1 \vert k} Q_T \bar{x}^*_{k+1 \vert k} \\
&\geq \ell(\bar{x}^*_{k \vert k}, \bar{u}^*_{k \vert k}) + (\bar{x}^{*}_{k+1 \vert k})^\top (Q + K^\top R K) \bar{x}^*_{k+1 \vert k} + (\bar{x}^{*}_{k+1 \vert k})^\top (\hat{A} + \hat{B}K)^\top Q_T (\hat{A} + \hat{B} K) \bar{x}^*_{k+1 \vert k} \\
& = \ell(\bar{x}^*_{k \vert k}, \bar{u}^*_{k \vert k}) + q(\bar{x}^*_{k+1 \vert k})
\end{aligned}
\end{equation*}
where $q(x)$ is defined as a positive definite quadratic function $q(x) = x^\top(Q + K^\top R K + (\hat{A} + \hat{B}K)^\top Q_T (\hat{A} + \hat{B} K))x$ and we denote its Lipschitz constant with respect to $\lVert \cdot \rVert_\infty$ over the compact set $\mathcal{R}$ as $L_q$. The terms $\bar{x}_{k+1\vert k}^*$ and $\bar{u}_{k+1 \vert k}^*$ denote the nominal predicted state and control input at time $k+1$ under the optimal controller synthesized at time $k$ by solving~\eqref{eq:tightening}. Since $T_k = 1$, we have $x(k+1) \in \mathcal{X}_T$ and the local controller $u = Kx$ is a feasible solution for the robust OCP inner approximation~\eqref{eq:tightening} with horizon $T = 1$ and $x_0 = x(k+1)$. Therefore, at time $k+1$, the optimal cost function of problem~\eqref{eq:tightening} satisfies $J^*_{T}(x(k+1)) \leq q(x(k+1)) = q(\bar{x}^*_{k+1 \vert k} + \eta(k)) \leq q(\bar{x}^*_{k+1 \vert k}) + L_q \lVert \eta(k) \rVert_\infty$ with $T =1$. By construction, we have that $J^*_{MPC}(x(k+1)) \leq J^*_{T}(x(k+1))$ for $T = 1$ and therefore
\begin{equation*}
\begin{aligned}
	J^*_{MPC}(x(k)) &\geq \ell(\bar{x}^*_{k \vert k}, \bar{u}^*_{k \vert k}) + q(\bar{x}^*_{k+1 \vert k}) \\
	& \geq \ell(\bar{x}^*_{k \vert k}, \bar{u}^*_{k \vert k}) + J^*_{MPC}(x(k+1)) - L_q\lVert \eta(k) \rVert_\infty \\
	& \geq \alpha_3(\lVert x(k)\rVert_\infty ) + J^*_{MPC}(x(k+1)) - L_q\lVert \eta(k) \rVert_\infty 
\end{aligned}
\end{equation*}
for some class-$\mathcal{K}_\infty$ function $\alpha_3(\cdot)$ which satisfies condition~\eqref{eq:ISS_Lyap_cond_2}.

\noindent \textbf{Case 2 $T_k \geq 2$}: First, we note that
\begin{equation}
\begin{aligned}
J^*_{MPC}(x(k)) &= \sum_{i=0}^{T_k-1} \ell(\bar{x}^*_{k+i \vert k}, \bar{u}^*_{k+i \vert k})) + (\bar{x}^*_{k+T_k \vert k})^\top Q_T \bar{x}^*_{k+ T_k \vert k} \\
&= \ell(\bar{x}^*_{k\vert k}, \bar{u}^*_{k \vert k}) + q(\bar{x}^*_{k+1 \vert k})
\end{aligned}
\end{equation}
where $q(\cdot)$ in this case is defined as the sum of the quadratic terms in the tail under the truncated controller~\eqref{eq:robust_policy_kp}, and the Lipschitz constant of $q(\cdot)$ with respect to $\lVert \cdot \rVert_\infty$ over the region $\mathcal{R}$ is denoted by $L_q$. By Theorem~\ref{thm:recursive_feasibility} on the recursive feasibility of the proposed MPC method, if $T_k \geq 2$ at time $k$, then at time $k+1$ the robust OCP~\eqref{eq:tightening} is feasible with horizon $T_k -1$ with the shifted controller~\eqref{eq:robust_policy_kp} as a feasible solution. Therefore, we have $J^*_{MPC}(x(k+1) \leq J^*_{T_k -1}(x(k+1)) \leq q(x(k+1))$ which leads to

\begin{equation}
\begin{aligned}
J^*_{MPC}(x(k+1)) & \leq q(x(k+1)) \\
&= q(\bar{x}^*_{k+1 \vert k} + \eta(k)) \\
& \leq q(\bar{x}^*_{k+1 \vert k} ) + L_q \lVert \eta(k) \rVert_\infty \\
& = J^*_{MPC}(x(k)) - \ell(\bar{x}^*_{k\vert k}, \bar{u}^*_{k \vert k}) + L_q \lVert \eta(k) \rVert_\infty \\
& \leq - \alpha_3(\lVert x(k) \rVert_\infty) +  L_q \lVert \eta(k) \rVert_\infty
\end{aligned}
\end{equation}
for some class-$\mathcal{K}_\infty$ function $\alpha_3(\cdot)$. Summarizing both cases, we conclude that $J^*_{MPC}(\cdot)$ is an ISS-Lyapunov function which shows that the closed-loop system~\eqref{eq:LTI_cl_dyn} is ISS.

\newpage
\appendix

\end{document}